\documentclass{aa}  

\usepackage{graphicx}
\usepackage{natbib} 
\usepackage{amsmath}
\usepackage{txfonts}
\begin{document} 

   \title{Reflection geometries in absorbed and unabsorbed AGN}

   \author{C. Panagiotou
          \and
          R. Walter
          }

   \institute{
              Astronomy Department, University of Geneva, Chemin d’Ecogia 16, 1290 Versoix, Switzerland\\
              \email{Christos.Panagiotou@unige.ch} 
             }

   \date{Received ***; accepted ***}

 
  \abstract
   {The hard X-ray emission of active galactic nuclei (AGN), and in particular, the reflection component, is shaped by the innermost and outer regions of the galactic nucleus.}
   {Our main goal is to investigate the variation of the Compton hump amongst a population of sources and correlate it with other spectral properties to constrain the source geometry. }
   {We studied the NuSTAR hard X-ray spectra of a sample of 83 AGN and performed a detailed spectral analysis of each of them. Based on their spectral shape, we divided the sample into five categories and also studied their stacked spectra. }
   {We found a stronger reflection in mildly obscured sources, which verifies the results reported in previous works. In addition, the reflection behaviour, and probably origin, varies with absorption. The accretion disc seems to be the main reflector in unabsorbed sources. A clumpy torus seems to produce most of the reflection in obscured sources. The filling factor of the clouds surrounding the active nucleus is a key parameter that drives the appearance of AGN. Finally, we found that the Fe line and the Compton hump are roughly correlated, as expected. }
  {}


   \maketitle
%

\section{Introduction}

Active galactic nuclei (AGN) are powered by the accretion of matter onto a supermassive black hole. The liberated gravitational energy heats the accretion disc, which emits thermal radiation. The bulk of the disc emission resides in the optical and UV part of the electromagnetic spectrum.

In addition to the optical and UV emission, AGN are luminous X-ray sources. The X-ray emission is believed to be produced by the thermal Comptonisation of disc photons by high-energy electrons. Variability and microlensing studies place the X-ray source, usually referred to as the corona, at only a few gravitational radii away from the black hole.

According to the AGN unification model, the central engine is surrounded by a donut-shaped dusty region, the so-called torus. This structure has been proposed to account for the observational differences between the two types of AGN, that is, the lack of broad emission lines in the optical spectrum of Type 2 objects \citep{1993ARA&A..31..473A}. If it lies on our line of sight, the dusty region prevents the broad lines from reaching us and also absorbs part of the X-ray emission.

Initial studies proposed a homogeneous torus extending up to 10 pc away from the black hole and aligned with the outer part of the accretion disc. The torus homogeneity has been challenged by contemporary data, and recent studies favor the existence of a clumpy torus. Clumpy torus models \citep[e.g.][]{2008ApJ...685..147N,2008ApJ...685..160N} have been successful in reproducing the observed infrared spectral energy distribution (SED) of AGN \citep[e.g.][]{2013ApJ...764..159L}. The clumpiness of the torus has also been indirectly verified by the observation of temporal X-ray eclipses in several AGN \citep{2014MNRAS.439.1403M}, which were attributed to the passage of individual absorbing clouds through our line of sight. Recent reviews of the torus physics and the absorption in AGN are given by \cite{2015ARA&A..53..365N} and \cite{2017NatAs...1..679R}.

In general, the AGN X-ray spectrum above 3 keV can be decomposed into three constituents: i) the primary Comptonisation emission, which is described well by a cut-off power law, ii) the Fe K$\alpha$ emission line at 6.4 keV, a ubiquitous spectral feature of AGN, and iii) the Compton hump, an excess emission in comparison to the power law at energies higher than 10 keV, which peaks at $\sim$ 30 keV. The Fe emission line has been suggested to be produced by absorption of the primary X-rays in the upper layers of the disc. In a handful of AGN, reverberation studies have confirmed that the line is formed in the inner parts of the accretion disc \citep[][and references therein]{2016MNRAS.462..511K}. The Compton hump is produced by the scattering of high-energy continuum photons by the surrounding material. It is still highly debated whether this reflection takes place on the disc or in the clouds of the torus. 

The investigation of the hard X-ray spectrum of AGN has been the target of several studies over the past decade. \cite{2011A&A...532A.102R} analysed the average spectra obtained with the International Gamma-Ray Astrophysics Laboratory (INTEGRAL)  \citep{2003A&A...411L...1W} of 165 nearby Seyfert galaxies and showed that the reflection emission is stronger in mildly obscured ($23<log(N_H)<24$) than in unobscured sources. Recently, \cite{2016A&A...590A..49E} obtained a similar result using stacked spectra obtained with the Burst Alert Telescope (BAT) \citep{2005SSRv..120..143B} of the same sample. A straightforward explanation of this result within the unification model is difficult because obscured sources are observed through higher inclination angles (i.e. closer to edge-on); in this case, the disc or torus reflection is expected to be weak. This result also has important implications for the fraction of Compton-thick (CT) AGN because it constrains their contribution to the cosmic X-ray background (CXB) flux to less than 6\%.

The Nuclear Spectroscopic Telescope Array (NuSTAR) \citep{2013ApJ...770..103H} is the first telescope that focuses X-rays above 10 keV. Its unprecedented sensitivity to photons with energy 3-78 keV allows for a simultaneous observation of the absorbed and reflected part of the AGN spectra and has provided new insights into the AGN physics. Recently, \cite{2017ApJ...849...57D} studied the average spectrum of 182 AGN detected by NuSTAR in the Extended Chandra Deep Field-South, Cosmic Evolution Survey, Extended Groth Strip, and the serendipitous survey fields. They found an average power-law slope of $\Gamma = 1.65$ and a mean reflection strength of $R \simeq 0.5$, where $R$ is a measure of the Compton hump intensity with respect to the continuum (Section \ref{Sec:spectra_fit}). It was also shown that $R$ decreases with the 10-40 keV luminosity.

In a companion paper, \cite{2018ApJ...854...33Z} studied the individual 0.5-24 keV spectra of the brightest hard X-ray sources detected in these surveys. They found a median reflection strength of $R = 0.43$ and also confirmed the anti-correlation between reflection and X-ray luminosity.

In this work, we study archived NuSTAR observations of a sample of nearby Seyfert galaxies. Our main aim is investigating their X-ray spectral properties and the evolution of the Compton hump strength in the different classes of AGN.

This paper is organised as follows. Section 2 describes the AGN sample and the data reduction. We present in Section 3 the spectral classification of the sources and describe the analysis in Section 4. Section 5 consists of a discussion of our main results, which are finally summarised in Section 6.

\section{Data sample and reduction}

\subsection{The whole sample}

We studied the sources contained in the final sample of \cite{2011A&A...532A.102R}, which were observed by NuSTAR until August 2017. At that time, 87 of their 165 sources had available public data in the NuSTAR archive. Our sample consists of 7 narrow-line Seyfert 1 (NLS1), 13 Seyfert 1, 14 Seyfert 1.5, 43 Seyfert 2, and 10 CT galaxies. The redshift of the individual sources ranges from 0.001 to 0.164. Most objects have been observed as part of the extragalactic survey of NuSTAR. Details for the different objects and their observations are given in Table \ref{tab:sources_info}.

The data reduction was performed using the NuSTAR Data Analysis Software (NuSTARDAS) version 1.8.0, which is part of the HEASoft package version 6.22. The raw events were calibrated and screened with the software tool \textit{nupipeline}. Events that occurred during the passage of the telescope through the South Atlantic Anomaly area were removed when necessary. For each observation, we visually examined the deduced sky image and defined a circular source region, centred on the corresponding object, with an optimal radius to maximise the signal-to-noise ratio. Because the brightness among the considered sources varies, the extraction radius is different for each object. The background was extracted from an annular region around the source; the inner radius of the annulus was 30 arcseconds larger than the radius of the source region in order to avoid source contamination. The background and source spectra, as well as the response files, were acquired using the \textit{nuproducts} tool. Several sources were observed more than once (see Table \ref{tab:sources_info}). We computed the time-averaged spectrum for each of these objects using the tool \textit{addspec}. The source spectra were binned so that each spectral bin contained at least 25 source counts. 

\subsection{Remarks on specific sources}

\textit{IGR J20286+2544}: This source, originally observed by INTEGRAL, has been found to be a dual AGN, containing NGC 6921 and MCG +04-48-002 \citep{2016ApJ...824L...4K}. Because NGC 6921 is classified as a low-ionisation nuclear emission-line region (LINER), we chose to study only MCG +04-48-002. We used a circular background region on this source.

\textit{NGC 1068}: This source has a complex spectrum that does not fit into any of the classes defined in Section \ref{classif}. We excluded it from our sample. A NuSTAR study of NGC 1068 has been conducted by \cite{2015ApJ...812..116B}. This CT AGN was found to have a reflection-dominated spectrum, which required the existence of a multi-component reflector to be well reproduced. Combining Chandra and NuSTAR data, these authors also showed that almost 30\% of the Fe K$\alpha$ emission originates from a region more than 140 pc away for the black hole, and thus well outside the torus environment. 

\textit{IGR J07565-4139, IGR J18259-0706}: These sources were excluded from the subsequent analysis. Because of their faintness, the model parameters could not be constrained.  

\textit{MCG-05-23-016}: The mean spectrum of this source is not fitted well by the simple model used in this work. To avoid dealing with specific characteristics of individual sources, we excluded this source from our sample. The NuSTAR observations of MCG-05-23-016 have been studied by \cite{2017ApJ...836....2Z}. They found that this Compton-thin ($N_H \sim 1.4 \cdot 10^{22} cm^{-2}$) source show features of relativistic reflection from the inner disc, while the system geometry remained constant over an observing period of two years.

\section{Classification}
\label{classif}

A preliminary examination of the spectra indicated that the sources could be divided into different groups. We decided to categorise the sources based on their X-ray spectral shape instead of using the conventional optical classification. Our aim was not to produce a robust X-ray classification, but rather to look for similarities among the sources that look similar, and to explore the differences between the various groups.

\begin{table}
        \centering
        \caption{Classification of sources }
        \label{tab:classification}
        \begin{tabular}{lccccccr} 
                \hline
                          &NLS1       &  Seyf 1   &Seyf 1.5     &Seyf 2                &CT       & \vline    & Total \\
                          &           &              &                &                &         & \vline \\
                \hline
                Class 1   & 1         & 4         & 3              & 3                & --       & \vline    & 11  \\
                Class 2   & 6         & 8         & 8              & 4                & --       & \vline    & 26  \\
                Class 3   & --        & --        & 1              & 14               & --       & \vline    & 15  \\
                Class 4   & --        & --        & 2              & 7                & 2        & \vline    & 11  \\
                Class 5   & --        & --        & --             & 13               & 7        & \vline    & 20  \\
                \hline
                Total     & 7           & 12      & 14             & 41               & 9        & \vline    & 83 \\
                \hline  
        \end{tabular}
        \tablefoot{Classes 1 to 5 correspond to the categories defined here. Each of them may include sources that are conventionally classified as NLS1, Seyfert 1 (Seyf 1), 2 (Seyf 2), and 1.5 (Seyf 1.5) or CT.}
\end{table}

\begin{figure}
  \centering
  \includegraphics[width=\linewidth,height=0.7\linewidth, clip]{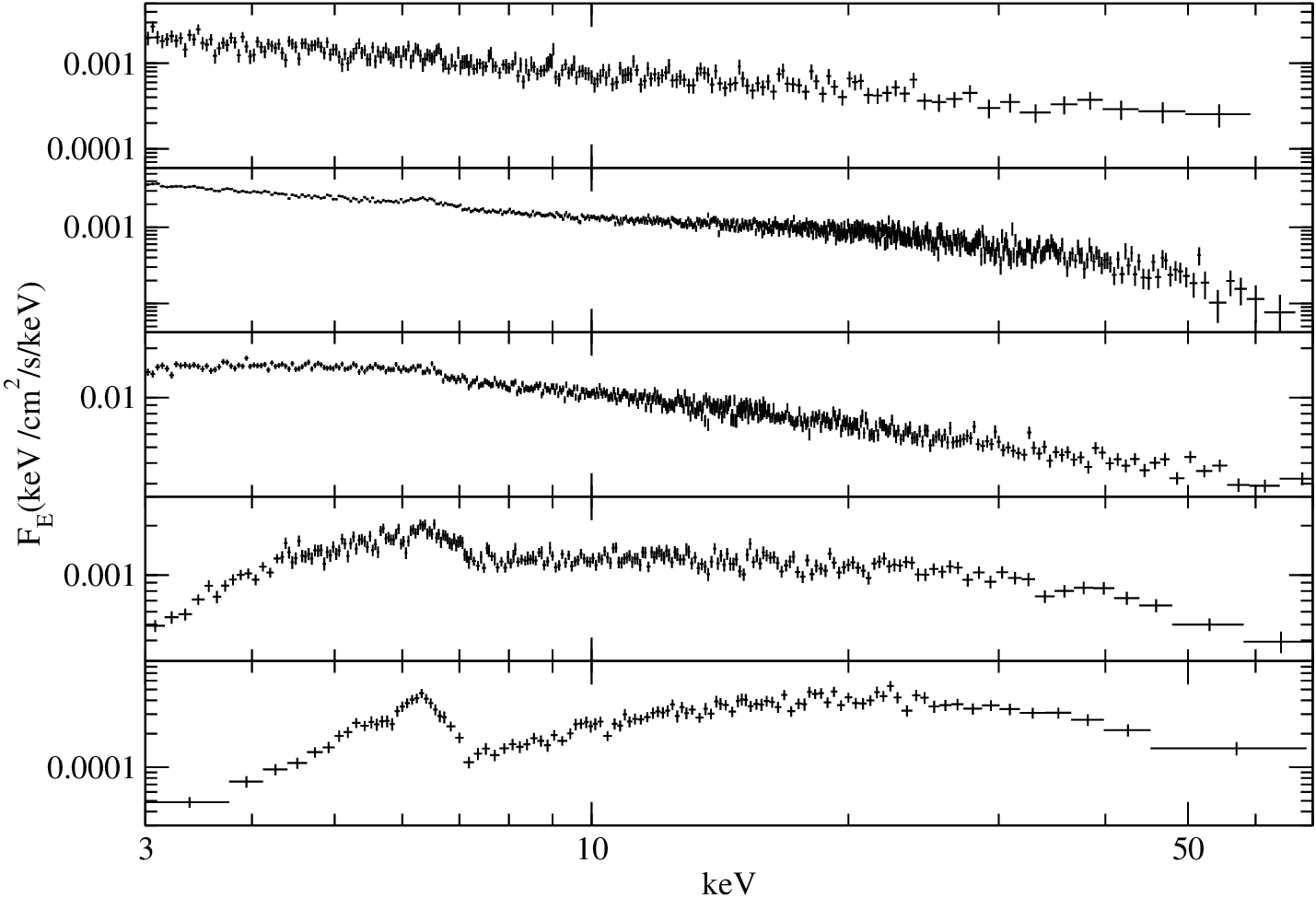}
  \caption[10]{Example of the spectra of the various classes. From top to bottom, the spectra correspond to the objects PG1501+106 (Class1), IGR J21277+5656 (Class 2), NGC 2110 (Class 3), NGC 6300 (Class 4), and MCG-03-34-064 (Class 5). }
  \label{fig:spec_classes}
\end{figure}

We thoroughly examined all the spectra visually and classified the sources into five categories. A typical example for the spectra of each class is shown in Fig. \ref{fig:spec_classes}. While the differences amongst the classes might be blurred to some extent, the common spectral characteristics within each category are the following. If we neglect the Fe line emission for the moment, the spectra in Class 1 seem to follow a simple power law, whereas Class 2 spectra slightly deviate from a power law in the energy range 10-40 keV. The spectra of Class 3 deviate from a power law at lower energies, below 5 keV, where the emission becomes flatter or even decreases. The emission of the sources in Class 4 seem to increase with energy at energies below 5 keV, while it remains roughly constant from 10 to $\sim$20 keV. Finally, the flux in Class 5 objects increases with energy both in the soft regime below 5 keV and at higher energies of 10 to 20 keV. The terms "decrease", "increase" and "flat" are used by reference to a spectrum in units of energy per unit time, area, and energy, as in Fig. \ref{fig:spec_classes}.

The classification of each source is listed in Table \ref{tab:sources_info}, and a summary with the number of sources in each class is given in Table \ref{tab:classification}. It is evident that our classification does not have a 1:1 correspondence with the conventional classification. 

\section{Spectral analysis}
\subsection{Classes 1-4}
\label{Sec:spectra_fit}

In this section we discuss the spectral results of the first four classes. Because they are more complex, Class 5 objects are considered separately in section \ref{sec:class_5}. The model fitting was performed using the XSPEC software \citep{1996ASPC..101...17A}, using the $\chi^2$ statistics. We also used the element abundances given by \cite{2009ARA&A..47..481A}. All the errors reported throughout the paper correspond to a 1$\sigma$ confidence interval, unless otherwise noted. For each source, we fitted the spectra obtained by the two NuSTAR detectors, FPMA and FPMB, accounting for a cross-normalisation, which is usually of the order of 5\% or lower.

We used a three-component model to fit the spectra. The main component, \texttt{pexrav} in XSPEC terminology, accounts for a power-law emission with an exponential high-energy cut-off plus a reflection hump \citep{1995MNRAS.273..837M}. This model simulates the reflection produced by a slab of infinite optical depth that is illuminated by the power-law continuum. Such a reflection is expected to resemble the reflection produced by an accretion disc well, while the power law describes the Comptonisation spectrum observed in AGN. The strength of the reflection with respect to the primary emission is given by the so-called reflection parameter $R$, which is an indicator of the reflector's covering factor. For an infinitely large disc and an isotropic source, $R$ should be equal to 1.

We also considered the \texttt{zphabs} model to account for photo-electric absorption in the AGN environment. We have ignored the absorption by the Galactic interstellar medium. The Galactic absorption column density towards all the sources in our sample is well below $10^{22}$ cm$^{-2}$ and therefore has no measurable effect on the spectra above 3 keV, where NuSTAR is sensitive.

Finally, we have included a Gaussian emission line at 6.4 keV to simulate the Fe K$\alpha$ line, which is prominent in many AGN. For simplicity, we kept the width of this line fixed at 0.05 keV when it could not be constrained by the fit. We further assumed solar abundance for the reflector and an inclination of $\text{cos } i = 0.45$ for all the sources. The free parameters to be minimised during the fit were the absorption column density, the power-law slope, the high-energy cut-off, the reflection strength, and the two normalization of the models, plus the cross-normalisation between the two instruments.

We used this model to fit all the spectra in classes 1-4. Most of the sources are well fitted, with a null hypothesis probability $P_{null}>5 \%$. We also reviewed the residuals of each fit, but no specific trend was prominent. A more complicated model is therefore not needed, except for two sources.

The quality of the fit for NGC 4051 was moderate with a $\chi^2$ of 1541 for 1270 degrees of freedom (dof). The examination of the residuals revealed the existence of an absorption line at $\sim$6.7 keV. Because NGC 4051 is well known to host a warm absorption outflow \citep[e.g.][]{2001ApJ...557....2C}, we decided to include a Gaussian absorption line component in our model as well. The new model improved the fit significantly, with $\chi^2/dof = 1484/1267$. The best-fit energy of the absorption line was E=$6.75_{-0.07}^{+0.06}$keV, while the remaining parameters are similar to the parameters that were obtained with the simpler model. We therefore kept this model for NGC 4051.

The best-fit model of IGR J19378-0617 suggested a very broad Fe line, with a width of $\sigma\sim1$kev. Modelling the iron line as a narrow line required a second emission line at $\sim6.9$keV, possibly produced by ionised Fe or by an outflow. Because the model with two emission lines provides more natural results for the width of the Fe K-alpha emission, we decided to use this model.

Moreover, three sources in Class 4, SWIFT J1930.5+3414, Mrk 477, and Mrk 6, had extremely hard spectra, with best-fit slopes of $\Gamma < 1.4$. Low index values like this are uncommon in AGN. It is more likely that they are just an artefact of the fit because the shape of the spectrum from 10 to 50 keV may also be well fitted by a hard power law with a small high-energy cut-off (38-48 keV) and no reflection, instead of a softer power law with a higher reflection strength. Therefore, we repeated the fit for these objects, but this time kept the slope fixed to the mean value of Class 4, that is, $\Gamma = 1.76$. The new fit was also statistically accepted, with a null hypothesis probability $P_{null}>10 \%$ for all the three spectra, and resulted in higher values for the absorption, the reflection, and the cutoff energies. These sources are denoted by crosses in the middle panel of Fig. 4.

\begin{figure}
  \centering
  \includegraphics[width=\linewidth,height=0.9\linewidth, clip]{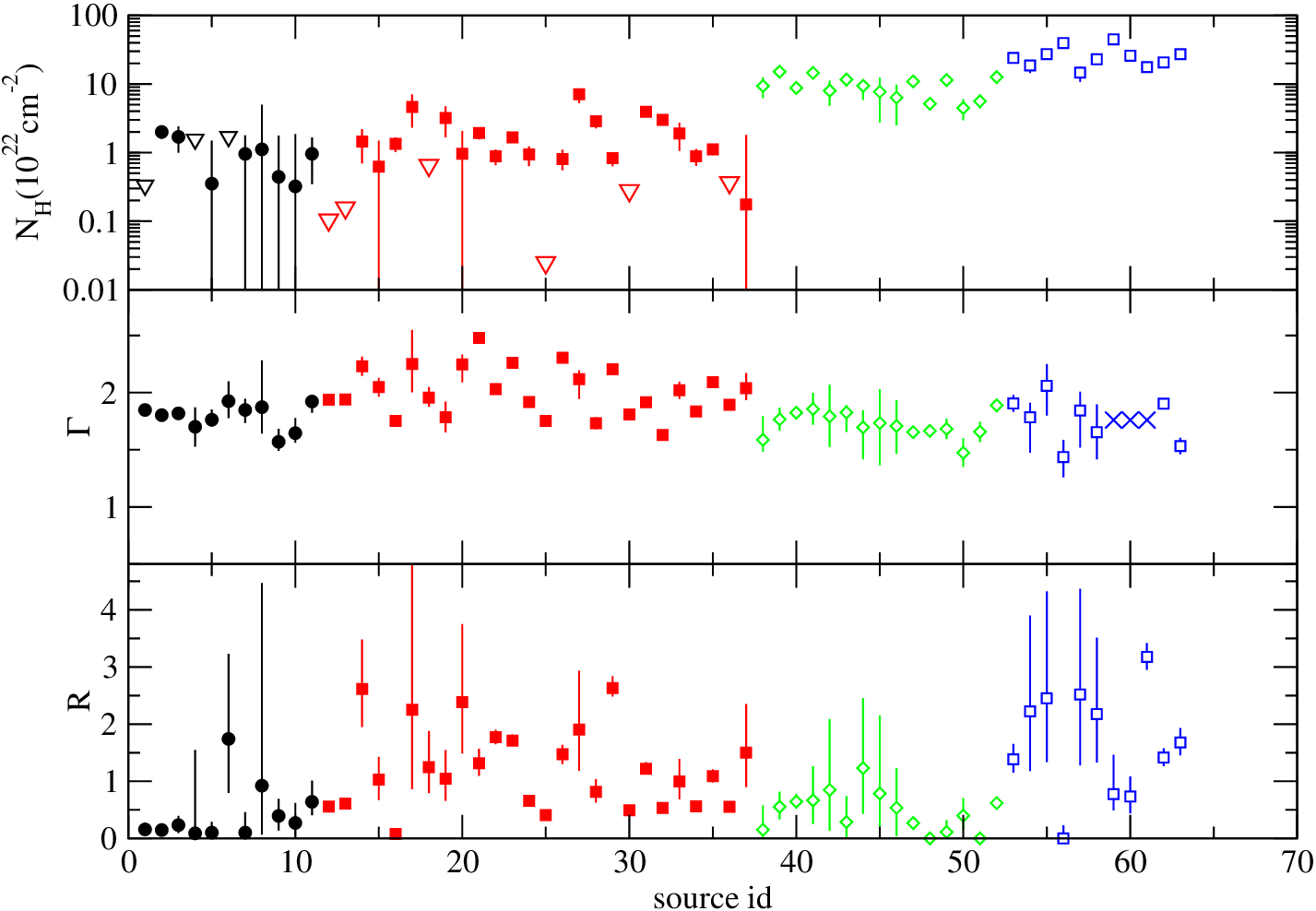}
  \caption[10]{Best-fit values of absorption column density (top), power-law slope (middle), and reflection strength (bottom panel). Class 1 objects are denoted by filled black circles and Class 2 objects by filled red squares. The open green diamonds indicate Class 3 sources and the open blue squares correspond to Class 4 objects. The open downward triangles in the top panel denote the 1$\sigma$ upper limits of these values. The crosses in the middle panel denote the sources that were fitted with a fixed $\Gamma$ value (see text for details).  }
  \label{fig:best_fit}
\end{figure}

The best-fit results are listed in Table \ref{tab:best_fit} and are plotted in Fig. \ref{fig:best_fit}. The top panel shows the best-fit absorption column density, the middle panel the best-fit spectral slope, and the deduced reflection values are plotted in the bottom panel. The different colours and symbols indicate the different classes as described in figure caption.

The absorption increases from Class 1 to 4, although the difference between the first two categories is not highly significant. Moreover, the sources of Class 2 on average have a harder spectrum than the rest. More precisely, the mean value of the slope is $1.79\pm0.03$, $2.01\pm0.04$, $1.72\pm0.03,$ and $1.76\pm0.05$ for the respective classes. In total, the mean value for the whole sample is $\langle \Gamma \rangle = 1.86 \pm 0.03$. This value is consistent with the value obtained by \cite{2018ApJ...854...33Z}, that is, $\langle \Gamma \rangle = 1.89 \pm 0.26 $, who have studied the NuSTAR spectra of more distant AGN.
Finally, the reflection emission seems to be stronger in the objects of Class 2 and 4. The mean reflection strength for all the sources is $\langle R \rangle = 0.99 \pm 0.10$, 
which is consistent with the interquartile range 0.06-1.50 reported by \cite{2018ApJ...854...33Z}. Our estimation of $\langle R \rangle$ differs from the value $ 0.5 ^{+0.15}_{-0.14} $ obtained by \cite{2017ApJ...849...57D}. The reason might be that they considered stacked spectra of faint sources and a fixed spectral slope.

\subsection{Class 5}
\label{sec:class_5}

The spectral shape of Class 5 objects points towards a reflection-dominated spectrum for these sources. We decided to fit their spectra with a more physical model. Our main aim was to determine whether a model of a compact homogeneous torus is good enough to reproduce the observed spectral densities or if a more complicated model is required by the data. Below we describe the different models we considered and the statistical tools we used to compare them. Our results are summarised in Table \ref{tab:class5_final}.

The spectral model we used is called MYTorus \citep{2009MNRAS.397.1549M, 2012MNRAS.423.3360Y}. This model assumes a geometry of a central X-ray source surrounded by a torus with a half-opening angle of 60 degrees. It consists of three separate components: i) the torus absorbed (including Compton scattering) primary power-law (called MYTZ), ii) the scattered emission (MYTS), which is produced by the reflection of the primary radiation in the torus, and iii) the iron Fe K$\alpha$ and K$\beta$ lines (MYTL), which are also presumed to be created by the reflection by the torus. The free parameters of this model are the power-law slope and normalisation of the primary continuum, the equatorial column density of the torus, and the inclination angle of the system.

Initially, we fitted the spectra with the so-called coupled version of the model. In this configuration, the scattered emissions and the primary continuum are coupled with a relative strength of one, which corresponds to the absorption and reflection being the result of a compact homogeneous torus. We found that the best-fit inclination angle was larger than 60 degrees for all the sources. In other words, all the spectra require the absorption of the primary X-ray continuum, which suggests that we observe these sources through the torus. 

Nearly half of the sources, that is, 13 out of 20, are well fitted with a null hypothesis probability of $P_{null}>5\%$. This model, however, provided a rather moderate fit for the rest of the sources and is insufficient to explain their emission. We noticed that for most sources, the inclination was found to be between 60 and 66 degrees. Because for a given column density the reflection strength is inversely proportional to the inclination angle, the best-fit model tries to combine an absorbed power-law spectrum with a significant contribution from reflection. This is indicative of a reflection-dominated spectrum.

We therefore proceeded to fit the spectra with the decoupled mode of MYTorus. Now, the primary and reflected component of the model are treated semi-indepedently; the slope and the column density are common in the three components, but the cross-normalisations between MYTS and MYTZ and between MYTL and MYTZ are left free to be minimized. The viewing angle for the primary continuum is fixed to be 90 degrees, that is, we observe the central source through the highest density region of the torus, while the MYTL and MYTS are observed in a face-on configuration. This model approximates the effects of a clumpy torus. 

The new model provides a fit that is statistically better or equivalently good to the fit we obtained with the coupled model (Table \ref{tab:class5_compar}). In order to quantify the difference in the quality of the fit between the two models, we calculated the Akaike information criterion, AIC \citep{Akaike1998,doi:10.1080/03610927808827599}:

\begin{eqnarray}
  \centering
      AIC_{c} = 2k - 2C_{L} + \chi^2 + \frac{2k(k+1)}{N-k-1},
\end{eqnarray}

\noindent where k is the number of free model parameters (five for the coupled version and six for the decoupled one), N is the number of data points, and $C_L$ is the likelihood of the true (and unknown) hypothetical model. This value is the same regardless of the used model. We also estimated the so-called evidence ratio, which for two models A and B being fit to the same data set, with $AIC_{c,B} < AIC_{c,A}$, is given by

\begin{eqnarray}
  \centering
      \epsilon_{A} = e^{-[AIC_{c,A} - AIC_{c,B}] / 2} 
\end{eqnarray}

\noindent The evidence ratio gives the relative probability between the two models, or in other words, it measures the likelihood that model B would provide such a better fit under the assumption that model A is the true model. If this likelihood is high, then the two models provide a statistically equivalent fit, and the simplest model is to be preferred. A detailed discussion of the interpretation of AIC and $\epsilon_{A}$ using astrophysical data has been given by \cite{2016MNRAS.461.1642E}.

\begin{table*}
        \centering
        \caption{Best-fit $\chi^2$ of the different models in Class 5 and their comparison. }
        \label{tab:class5_compar}
        \begin{tabular}{lrccccccc} 
                \hline
                Source             &  Source   & Coupled MYTorus     & Decoupled MYTorus    &$\epsilon$ (\%)      & {fixed $\Gamma$}    & F-test        & Model     \\
                 name              &  id       &  $\chi^2$/dof       & $\chi^2$/dof         &                     & $\chi^2$/dof        & $P_{null}$ (\%) &   \\
                \hline
   IGR J00254+6822       &64        &   48.9 / 42                   &     43.9 / 41        &30.8                 &   56.9 / 43         & 1.2         &CMwF        \\      
   IGR J02501+5440       & 65        &     9.7 / 21             &       8.1 / 20      &39.9                 &    9.7 / 22                 & 100.0             &CMwF     \\   
   SWIFT J0453.4+0404    & 66        &   889.4 / 774            &  774.9 / 773      &$4\cdot10^{-23}$      &  795.2 / 774        & $8\cdot10^{-4}$                  &DM   \\              
   SWIFT J0601.9-8636    & 67        &    62.9 / 44             &   64.5 / 43        &12.1                 &  106.9 / 45         & $2\cdot10^{-4}$                 &CM     \\             
   SWIFT J1009.3-4250    & 68        &   104.7 / 106            &   94.6 / 105      &2.0           &  106.9 / 107            & 13.9                         &CMwF   \\            
   IGR J13091+1137       & 69        &   277.9 / 231            &  263.4 / 230      &0.2           &  277.9 / 231            & 0.05                          &DM  \\              
   IGR J14175-4641       & 70        &    61.2 / 54             &   47.8 / 53        &0.4                  &   48.3 / 54         & 46.0                            &DMwF   \\             
   IGR J16351-5806       & 71        &    78.3 / 71             &   73.6 / 70        &31.8                 &   78.7 / 72         & 54.9                           &CMwF    \\             
   IGR J20286+2544       & 72        &    55.9 / 61             &   52.3 / 60        &55.9                 &   55.9 / 62               & 100.0                     &CMwF   \\             
   LEDA 15023            & 73        &   122.9 / 95            &  116.5 / 94       &12.4                 &  128.7 / 96                  & 3.7                       &CMwF   \\             
   LEDA 96373            & 74        &   118.5 / 100           &  104.3 / 99       &0.3                  &  110.4 / 100       & 1.8                          &DMwF   \\            
   MCG -03-34-064        & 75        &   783.2 / 567           &  643.5 / 566      &$10^{-28}$           &  713.7 / 567       & $2\cdot10^{-12}$             &DM   \\              
   Mrk 3                 & 76        &  5222.4 / 2207          & 2574.4 / 2206     &0.0                  & 2757.4 / 2207      & $8\cdot10^{-33}$               &DM   \\            
   NGC 1194              & 77        &   114.0 / 124           &  110.7 / 123      &58.6                 &  154.9 / 125           & $8\cdot10^{-8}$                  &CM   \\               
   NGC 3281              & 78        &   181.0 / 204           &  178.1 / 203      &67.1                 &  218.3 / 205       & $7\cdot10^{-8}$              &CM   \\              
   NGC 4507              & 79        &  2549.3 / 1674          & 1864.4 / 1673     &$5\cdot10^{-147}$    & 1940.7 / 1674      & $3\cdot10^{-14}$               &DM   \\            
   NGC 4941              & 80        &    46.2 / 35            &   43.8 / 34       &84.2                 &   48.5 / 36                  & 19.5                      &CMwF   \\             
   NGC 5643              & 81        &   126.5 / 111           &  127.9 / 110      &15.8                 &  286.1 / 112       & $2\cdot10^{-19}$             &CM   \\              
   NGC 5728              & 82        &   270.6 / 279           &  259.5 / 278      &1.0                  &  284.0 / 280       & 0.02                         &CM   \\              
   NGC 788               & 83        &   114.8 / 123           &   96.3 / 122      &$2.9\cdot10^{-2}$    &   99.1 / 123       & 6.2                          &DMwF   \\            
                \hline  
        \end{tabular}
        \tablefoot{The last column lists the model used for each source: coupled MYtorus (CM), decoupled MYtorus (DM), coupled MYtorus with fixed slope (CMwF), and decoupled MYtorus with fixed slope (DMwF).}
        \vspace{1cm}
\end{table*}

\begin{table*}
        \centering
        \caption{Best-fit parameter values of the different models in Class 5.}
        \label{tab:class5_fit}
        \begin{tabular}{ccccccccc} 
                \hline
                  Source        & \multicolumn{3}{c}{Coupled MYTorus}     & \multicolumn{2}{c}{Decoupled MYTorus}                 & fixed $\Gamma$    \\
                  id   &  $\Gamma$                &$N_H (10^{24}$ cm$^{-2})$  &Inclination (Degrees)      &  $\Gamma$                &$N_H (10^{24}$ cm$^{-2})$     &$N_H (10^{24}$ cm$^{-2})$     \\
                \hline
    64   &  $2.26 \pm 0.07$                 & $8.69_{-1.23}^{+1.25}$    & $60.81_{-0.19}^{+0.58}$        &        $1.58_{-0.18}^{+0.21}$     &    $1.35_{-0.26}^{+0.44}$      &   $6.75_{-0.55}^{+0.67}$         \\
    65   &  $1.81_{-0.26}^{+0.31}$    & $2.30_{-0.41}^{+0.40}$    & $74.58_{-7.42}^{+5.07}$         &  $1.89 \pm 0.30$           & $2.67_{-0.63}^{+0.71}$       & $2.29_{-0.32}^{+0.34}$         \\
    66   &  $2.33 \pm 0.01$           & $6.95_{-0.12}^{+0.14}$    & $60.78_{-0.06}^{+0.04}$       &  $1.97 \pm 0.03$           & $1.69_{-0.08}^{+0.09}$       & $1.36 \pm 0.02$                \\
    67   &  $2.14_{-0.05}^{+0.11}$    & $10.00_{-2.38}$           & $65.19_{-3.38}^{+5.84}$        &  $2.36_{-0.06}^{+0.11}$    & $9.99_{-4.93}^{+0.01}$       & $10.00_{-0.17}$         \\
    68   &  $1.51_{-0.08}^{+0.09}$    & $0.53_{-0.02}^{+0.08}$    & $87.62_{-11.23}^{2.38}$       &  $1.67 \pm 0.11$           & $0.74_{-0.10}^{+0.11}$       & $3.13_{-0.12}^{+0.16}$         \\
    69   &  $1.46_{-0.03}^{+0.02}$    & $1.95_{-0.06}^{+0.08}$    & $60.70_{-0.09}^{0.16}$       &  $1.40^{+0.12}$            & $0.57 \pm 0.02$              & $1.05 \pm 0.03$                \\
    70   &  $1.81_{-0.11}^{+0.18}$    & $2.86_{-1.09}^{+0.42}$    & $61.21_{-0.93}^{+1.80}$       &  $1.92_{-0.18}^{+0.17}$    & $1.45_{-0.27}^{+0.45}$       & $1.32_{-0.16}^{+0.18}$         \\
    71   &  $1.94_{-0.23}^{+0.17}$    & $3.89_{-0.26}^{+0.44}$    & $72.84_{-12.08}^{+4.58}$       &  $2.02_{-0.19}^{+0.17}$    & $4.67_{-0.53}^{+0.84}$       & $4.25_{-0.42}^{+3.58} $         \\
    72   &  $1.84 \pm 0.05$           & $6.91_{-0.41}^{+0.51}$    & $60.48_{-0.24}^{+1.43}$       &  $1.61_{-0.21}^{+0.18}$    & $1.12_{-0.27}^{+0.26}$       & $8.03_{-0.49}^{+0.58}$         \\
    73   &  $1.51 \pm 0.04$           & $2.01_{-0.59}^{+0.21}$    & $61.04_{-0.22}^{+1.01}$       &  $1.49_{-0.09}^{+0.16}$    & $0.73_{-0.09}^{+0.22}$       & $3.96_{-0.17}^{+0.20}$                \\
    74   &  $2.57_{-0.04}^{+0.03}$    & $9.03_{-0.50}^{+0.62}$    & $60.97_{-0.16}^{+0.29}$       &  $1.40^{+0.07}$            & $0.62 \pm 0.07$              & $1.14_{-0.06}^{+0.07}$         \\
    75   &  $2.42_{-0.01}^{+0.02}$    & $8.85_{-0.22}^{+0.23}$    & $60.82_{-0.05}^{+0.06}$       &  $2.35 \pm 0.02$           & $3.35 \pm 0.17$              & $1.35 \pm 0.03$                \\
    76   &  $1.663_{-0.004}^{+0.002}$ & $2.002_{-0.004}^{+0.009}$ & $61.25_{-0.02}^{+0.05}$       &  $1.673_{-0.008}^{+0.005}$  & $1.000_{-0.011}^{+0.003}$   & $1.097 \pm 0.007$              \\
    77   &  $2.16 \pm 0.03$           & $9.02_{-0.56}^{+0.61}$    & $60.82_{-0.12}^{+0.28}$       &  $1.91 \pm 0.08$           & $2.56_{0.31}^{+0.34}$        & $10.00_{-0.42}$         \\
    78   &  $2.03_{-0.02}^{+0.03}$    & $9.05_{-0.32}^{+0.60}$    & $61.19_{-0.30}^{+0.33}$       &  $2.07_{-0.05}^{+0.03}$    & $5.35_{-0.91}^{+0.76}$       & $10.00_{-0.12}$         \\
    79   &  $1.700_{-0.007}^{+0.004}$ & $2.02 \pm 0.01$           & $61.36_{-0.13}^{+0.05}$       &  $1.68 \pm 0.01$           & $0.95 \pm 0.02$              & $1.06 \pm 0.01$                \\
    80   &  $2.01_{-0.11}^{+0.10}$    & $4.87_{-2.15}^{+1.77}$    & $61.33_{-2.16}^{+3.97}$       &  $1.56_{-0.16}^{+0.24}$    & $1.43_{-0.32}^{+0.58}$       & $3.84_{-1.66}^{+0.70}$         \\
    81   &  $2.45_{-0.08}^{+0.10}$    & $9.00_{-2.62}^{+1.00}$    & $62.90_{-1.10}^{+5.44}$       &  $2.60_{-0.71}$            & $10.00_{-0.61}$              & $5.00_{-0.32}^{+0.36} $         \\
    82   &  $1.58 \pm 0.04$           & $2.00_{-0.51}^{+0.13}$    & $65.88_{-0.90}^{+8.00}$       &  $1.63 \pm 0.05 $          & $1.44_{-0.08}^{+0.09}$       & $1.68_{-0.08}^{+0.09} $         \\
    83   &  $1.58_{-0.04}^{+0.03}$    & $1.99_{-0.09}^{+0.22}$    & $61.10_{-2.83}^{+0.70}$       &  $1.62_{-0.19}^{0.11}$     & $1.06_{-0.25}^{+0.15}$       & $1.26_{-0.10}^{+0.09}$         \\
                \hline  
        \end{tabular}
        \tablefoot{The $N_H$ parameter of the used model is limited to values below $10^{25}$cm$^{-2}$.}
        \vspace{1cm}
\end{table*}

The estimated values of the evidence ratio for each source are listed in Table \ref{tab:class5_compar}. For the sources with $\epsilon > 1\%$, we assumed that the simpler model, that is, the coupled version of MYTorus, is to be preferred, while for the remaining sources, we used the decoupled model. We found that 8 out of 20 sources required the decoupled model. The best-fit parameter values for both models are given in Table \ref{tab:class5_fit}.

\begin{figure}
  \centering
  \includegraphics[width=\linewidth,height=0.9\linewidth, clip]{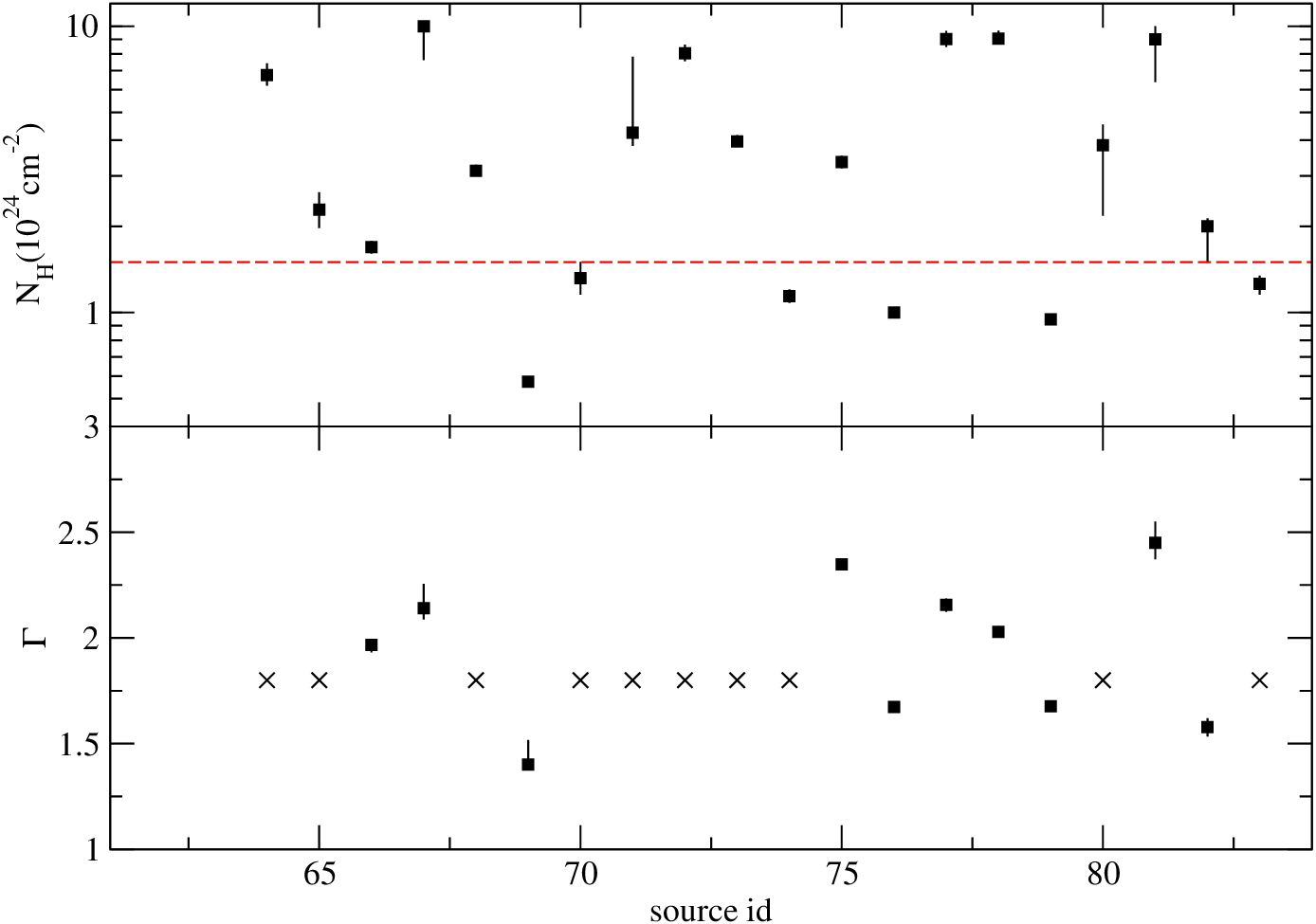}
  \caption[10]{Best-fit values of absorption column density (top) and power-law slope (bottom panel) for the reflection-dominated objects (Class 5). The red dashed line in the top panel denotes the boundary ($N_H = 1.5 \cdot 10^{24} cm^{-2}$) between CT and Compton-thin sources.}
  \label{fig:best_fit_class5}
\end{figure}

We observed that the best-fit $N_H$ and $\Gamma$ are correlated. This is probably an inherent correlation of the model or due to the low statistics of our spectra. We decided to rerun the fit for all the sources with the slope being fixed at the weighted mean best-fit value, $\Gamma=1.8$. 

We performed an F-test to decide if the model with a free $\Gamma$ provides a significantly better fit. The F-test is a statistical test that is used to compare the goodness of the fit between two models, with one model being nested to the other. It gives the probability, $P_{null}$, that both models fit the data similarly well. If $P_{null} \ll 1$, then the most complex model, that is, the model with more free parameters, should be preferred. In our case, a small $P_{null}$ would indicate that $\Gamma$ should be left free during the fit.

The probabilities calculated with the F-test are listed in Table \ref{tab:class5_compar}. We found that $P_{null}>1\%$ for 10 sources, which suggests that the model with a fixed $\Gamma$ is statistically preferable in these cases. For the remaining sources, we continued to use the models with a free $\Gamma$. 

All the models we considered seemed insufficient to provide a good fit for Mrk 3. A visual examination of the fit residuals revealed a higher emission than expected by the models in the energy range below 5 keV. This could be the result of a warm absorber in the AGN environment or just an artefact produced by adding the 11 individual spectra of this source. Because the response of NuSTAR is less well calibrated in the 3-5 keV energy range \citep{2015ApJS..220....8M}, we decided to ignore the data below 5 keV for this source. Thus, we obtained a statistically accepted fit, with $\chi^2/dof = 2202 / 2106$. The best-fit parameter values are close to those presented by \cite{2016MNRAS.460.1954G}, who have analysed part of the NuSTAR observations of Mrk 3.

To recapitulate, we initially fitted all the spectra with both the coupled and decoupled version of MYTorus. We compared the two models using the evidence ratio. After deciding which of the two should be used for each source, we checked whether the power-law slope was to be fixed during the fit using the F-test. The model we finally considered for each source is listed in the last column of Table \ref{tab:class5_final}. The final best-fit absorption column densities and photon indices are plotted in Fig. \ref{fig:best_fit_class5} and listed in Table \ref{tab:class5_final}. The crosses in this figure denote the sources that were fitted with a fixed slope. Seventy percent (14 out of 20) of the sources have a best-fit $N_H$ in the CT regime.

\begin{table*}
        \centering
        \caption{Final best-fit absorption column density and power-law slope for the objects of Class 5.}
        \label{tab:class5_final}
        \begin{tabular}{ccccccccc}
                \hline
                 Source id   &  $\Gamma$          &$N_H (10^{24}$ cm$^{-2})$    &$\chi^2$/dof                 & Model       \\
                \hline
    64   &  $1.8 \text{\small(f)}$      & $6.75_{-0.55}^{+0.67}$    &  56.9 / 43     &CMwF   \\
    65   &  $1.8 \text{\small(f)}$    & $2.29_{-0.32}^{+0.34}$    &   9.7 / 22     &CMwF     \\
    66   &  $1.97 \pm 0.03$           & $1.69_{-0.08}^{+0.09}$    & 774.9 / 773     &DM   \\
    67   &  $2.14_{-0.05}^{+0.11}$    & $10.00_{-2.38}$           &  62.9 / 44     &CM    \\
    68   &  $1.8 \text{\small(f)}$    & $3.13_{-0.12}^{+0.16}$    & 106.9 / 107     &CMwF   \\
    69   &  $1.40^{+0.12}$            & $0.57 \pm 0.02$           & 263.4 / 230      &DM  \\
    70   &  $1.8 \text{\small(f)}$    & $1.32_{-0.16}^{+0.18}$    &  48.3 / 54     &DMwF   \\
    71   &  $1.8 \text{\small(f)}$    & $4.25_{-0.42}^{+3.58}$    &  78.7 / 72    &CMwF    \\
    72   &  $1.8 \text{\small(f)}$    & $8.03_{-0.49}^{+0.58}$    &  55.9 / 62     &CMwF   \\
    73   &  $1.8 \text{\small(f)}$    & $3.96_{-0.17}^{+0.20}$    & 128.7 / 96     &CMwF   \\
    74   &  $1.8 \text{\small(f)}$    & $1.14_{-0.06}^{+0.07}$    & 110.4 / 100     &DMwF   \\
    75   &  $2.42_{-0.01}^{+0.02}$    & $1.35 \pm 0.03$           & 643.5 / 566     &DM   \\
    76   &  $1.663 \pm 0.008$         & $0.93 \pm 0.02$           & 2201.8 / 2106     &DM   \\
    77   &  $2.16 \pm 0.03$           & $9.02_{-0.56}^{+0.61}$    & 114.0 / 124     &CM   \\
    78   &  $2.03_{-0.02}^{+0.03}$    & $9.05_{-0.32}^{+0.60}$    & 181.0 / 204     &CM   \\
    79   &  $1.68 \pm 0.01$           & $0.95 \pm 0.02$           & 1864.4 / 1673     &DM   \\
    80   &  $1.8 \text{\small(f)}$    & $3.84_{-1.66}^{+0.70}$    &  48.5 / 36     &CMwF   \\
    81   &  $2.45_{-0.08}^{+0.10}$    & $9.00_{-2.62}^{+1.00}$    & 126.5 / 111     &CM   \\
    82   &  $1.58 \pm 0.04$           & $2.00_{-0.51}^{+0.13}$    & 270.6 / 279     &CM   \\
    83   &  $1.8 \text{\small(f)}$    & $1.26_{-0.10}^{+0.09}$    &  99.1 / 123     &DMwF   \\
                \hline  
        \end{tabular}
        \tablefoot{The last column lists the model used for each source as in Table \ref{tab:class5_compar}.}
        \vspace{1cm}
\end{table*}

\subsection{Composite spectra}

In the following, we computed the average spectrum of each class. Our goal is to determine whether the spectral differences observed in the individual sources are still evident when we stack their spectra and to compare our results with previous studies. 

In order to account for the differences in the sizes of the source regions we used to extract the individual spectra, we rescaled all the spectra to a common source region. We did that by multiplying each spectrum with the ratio of its ARF file to a reference ARF file. For each class we chose one of its objects to be the reference source. Then, we added all the deduced spectra using the \textit{mathpha} tool. We also took into account the differences in the exposure times in order to substract the proper amount of background. The stacked spectra were grouped so that every energy bin contained at least 1000 counts.

\begin{figure}
  \centering
  \includegraphics[width=\linewidth,clip]{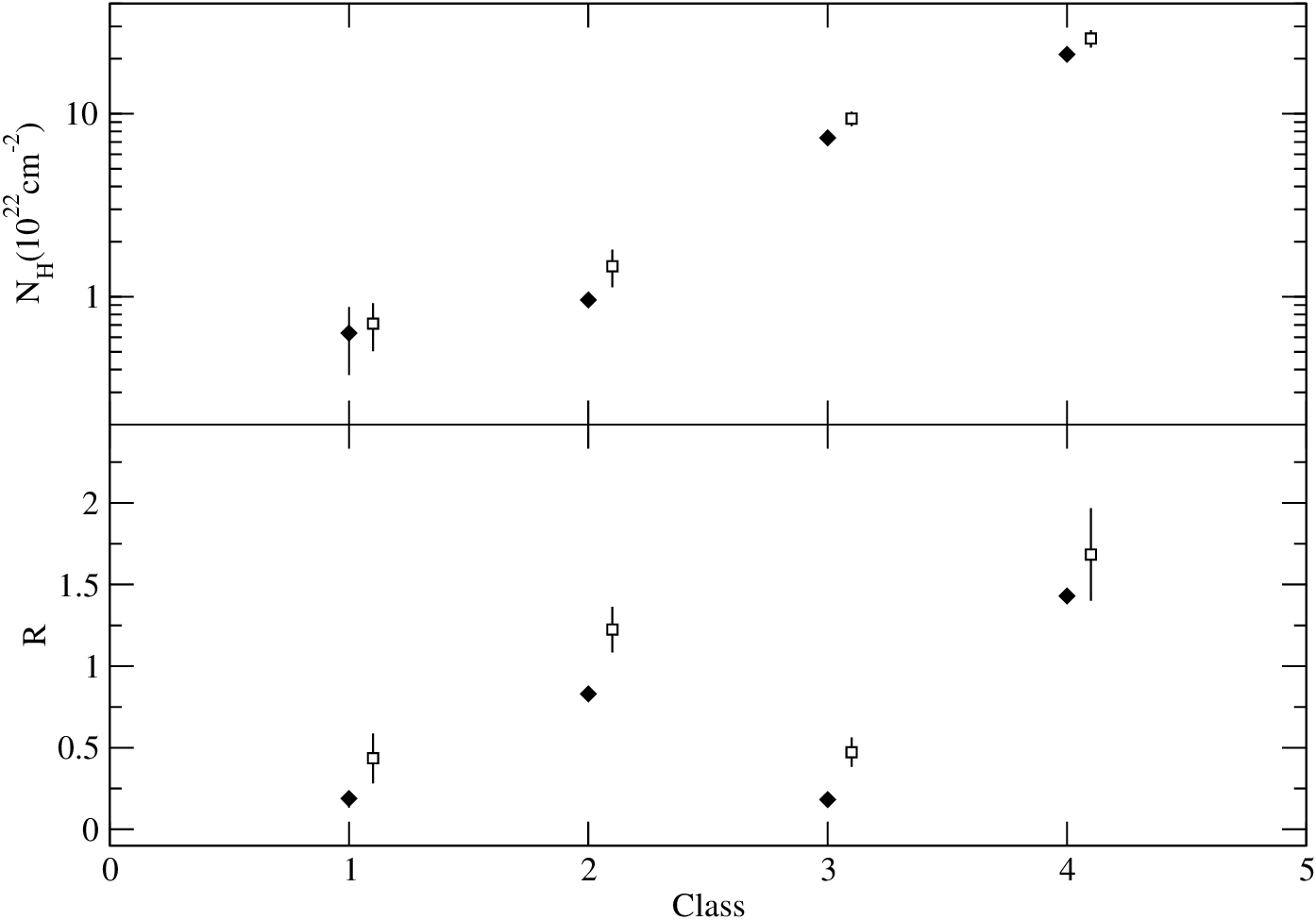}
  \caption[9]{Best-fit absorption column density (upper) and reflection strength (lower panel) for the different classes. The filled diamonds correspond to the values acquired when we fitted the composite spectra, and the open squares show the mean values of the best fits to the individual spectra. The arbitrary offset along the x-axis is to improve clarity.}
  \label{fig:spectra_class}
\end{figure}

After estimating the composite spectra, we then fitted them with the same model used for the individual sources. Our best-fit results for the absorption and reflection level are plotted in Fig. \ref{fig:spectra_class}. The best-fit values of the composite spectra are denoted by filled diamonds, while the open squares indicate the average values of the individual spectra best fits. 

The stacked spectrum values are mostly consistent within the errors to those that we obtained using only the individual spectra, and they reproduce the trends observed previously well; the absorption increases with the class, and the reflection is stronger in Classes 2 and 4. It should be pointed out that this agreement is not expected a priori because the stacking procedure may result in the creation of artefacts in the mean spectra.

\begin{figure}
  \centering
  \includegraphics[width=\linewidth,clip]{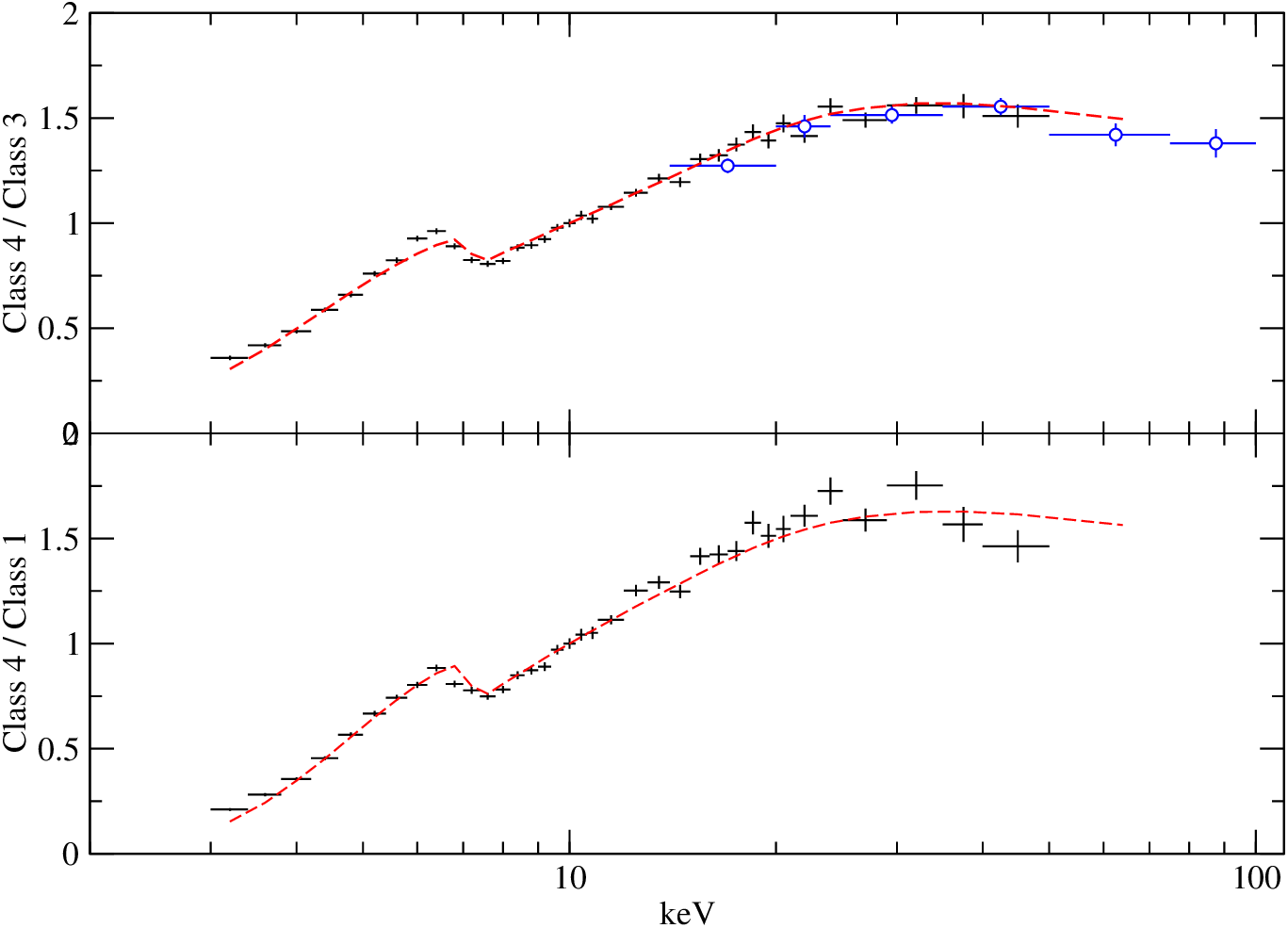}
  \caption[9]{Ratio of Class 4 spectrum to Class 3 (upper panel) and Class 1 (bottom). The blue circles correspond to the ratio of mildly obscured sources over lightly obscured objects using BAT stacked data \citep{2016A&A...590A..49E}. The dashed red line denotes the expected ratio between two theoretical spectra that differ only in the reflection and obscuration level.}
  \label{fig:count_ratios}
\end{figure}

The composite spectra can also be used to study differences amongst the various classes in a model-indepedent way. For example, the ratio between two spectra provides an indirect measurement for the relative reflection and absorption of one spectrum with respect to the other. Two such ratios are shown in Fig. \ref{fig:count_ratios}. The top panel plots the ratio of Class 4 to Class 3, while the bottom panel plots the ratio between Class 4 and 1. All the spectra were renormalised before the division so that they have the same value at 10 keV. 

Both ratios exhibit a similar curvature from 10 to 80 keV, which suggests an emission excess at these energies in Class 4 compared to Classes 3 and 1. This excess may naturally be attributed to a stronger Compton hump emission for Class 4. Below 10 keV, the two ratios increase with energy, and a Fe edge-like feature is observed at $\sim$7 keV. If we assume that the spectra in the different classes have similar photon index distributions, these characteristics imply that the absorption is on average higher in the Class 4 sources, as already inferred by the spectral analysis.

The red dashed curves in Fig. \ref{fig:count_ratios} denote the expected spectral ratios between two theoretical emissions of absorbed power law plus reflection. The two emissions have the same slope ($\Gamma=1.75$), differ only in the reflection and absorption levels, and were calculated using the models \texttt{phabs} and \texttt{pexrav} in XSPEC. A spectrum of $R=1.2$ and $N_H = 22\cdot 10^{22}$cm$^{-2}$ was divided by a spectrum of $R=0.2$ and $N_H = 10\cdot 10^{22}$cm$^{-2}$ (upper panel) and by a spectrum of $R=0.3$ and $N_H = 3\cdot 10^{22}$cm$^{-2}$ (lower panel). The general agreement between the two theoretical ratios and the data supports the results reported above.

The blue circles in the top panel of Fig. \ref{fig:count_ratios} denote the spectral ratio of mildly obscured sources ($23<log(N_H)<24$) versus lightly obscured ones ($21<log(N_H)<22$), as calculated by \cite{2016A&A...590A..49E} based on BAT data. The NuSTAR and BAT curves are not directly comparable because we have used a different classification. Nevertheless, both NuSTAR and BAT data point towards the same result. The high-energy spectral difference between mildly and less obscured sources is well explained by an increase in the reflection.

\section{Discussion}

\subsection{Reflection strength and the CXB}

The prominent reflection hump in AGN spectra may provide significant information about the geometry and the environment close to the nucleus. However, its strength remains one of the least constrained quantities in AGN, mainly because of the degeneracies of $R$ to other spectral parameters and because we lacked a highly sensitive hard X-ray instrument prior to NuSTAR. One of the main aims of this work was investigating $R$ among various AGN and its evolution with absorption using NuSTAR spectra.

We found that the reflection varies significantly amongst the different sources. More precisely, the higher $R$ values have been found for the objects of Classes 4 and 2, while the sources in Classes 1 and 3 exhibit low reflection levels, if any. It seems that in both the absorbed (Classes 3 and 4) and the unabsorbed or lightly absorbed (Classes 1 and 2) regime, the reflection strength takes a variety of values. This result is difficult to be explained within the unification model, if the reflector is the same in every AGN. For example, if the primary X-ray emission is reflected only by the disc, there is no simple explanation why the reflection in the absorbed sources is variable, or why Class 4 objects require high $R$ values. This is discussed further in the subsequent sections.

\cite{2011A&A...532A.102R} were the first to show that the reflection in mildly obscured sources ($23<log(N_H)<24$) is higher than expected. Using stacked INTEGRAL spectra, they found that the best-fit reflection of mildly obscured Seyfert 2 sources is $R=2.2$, remarkably higher than the value estimated for Seyfert 1, 1.5, and less obscured Seyfert 2 objects. This result was also obtained by \cite{2016A&A...590A..49E} with improved statistics, using stacked BAT spectra. 

The analysis of NuSTAR spectra of the same sample of sources verified this result. The mildly obscured sources belong mostly to Class 4, which is the class with the highest reflection value. More importantly, the sensitivity of NuSTAR enabled us to show that this result holds even when the individual spectra of each source are used, and we thus unambiguously confirmed that obscured AGN have strong reflection.

This result has an important implication for the CXB and the AGN population models as well. The CXB is the integrated emission of discrete AGN, and its spectrum peaks around 30keV. The reproduction of CXB requires knowing the spectral templates for the various AGN and also the demographics of the AGN population. The CXB has therefore been used to estimate the proportion of CT AGN in the universe. Nevertheless, this quantity remains largely unconstrained, with a huge variety of numbers found in the recent literature.

\cite{2016A&A...590A..49E} used BAT spectral templates to reproduce the CXB spectrum. They showed that the high reflection that is observed in the mildly obscured AGN causes these sources to contribute significantly to the CXB emission. This constrains the fraction of CT to the one effectively detected \citep[e.g.][]{2015ApJ...815L..13R, 2012MNRAS.423..702B}. Our results confirm that the high reflection is a common characteristic in the individual spectra of mildly obscured sources and that therefore no hidden CT population of AGN is required.

\subsection{Unabsorbed sources (Classes 1 and 2)}

\begin{figure}
  \centering
  \includegraphics[width=\linewidth,clip]{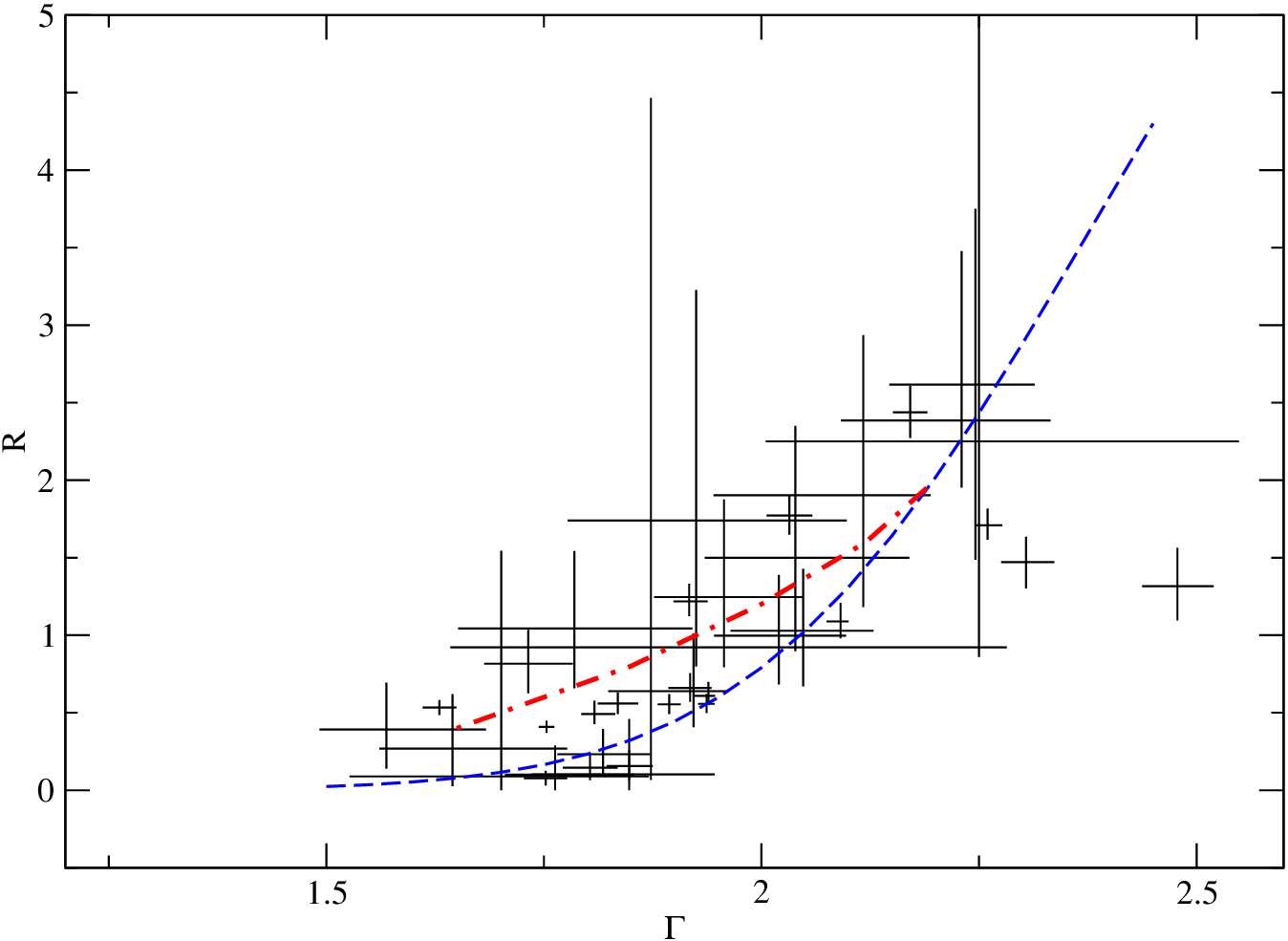}
  \caption[9]{Correlation between reflection and photon index in unabsorbed sources (Classes 1 and 2). The blue dashed line indicates the correlation expected as a result of the relative motion of the corona with respect to the disc (Beloborodov 1999). The red dot-dashed line represents the correlation expected when the covering factor of the disc is variable.}
  \label{fig:gamma_vs_refl}
\end{figure}

We then explored our best-fit results for any prominent correlations that would explain the reflection variability. At first, we concentrated on the less absorbed Classes 1 and 2. We found that the reflection is highly correlated with the power-law slope, as is clearly shown in Fig. \ref{fig:gamma_vs_refl}

However, these two quantities are intrinsically correlated in the spectral model. We ran extensive simulations to assess whether the observed correlation is real. We used the NuSTAR responses and the \texttt{pexrav} model to produce fake spectra of the same $R$ and $\Gamma$, and with exposure time and flux close to the mean value of the spectra of our sources. We then fitted the fake spectra and investigated the $R-\Gamma$ correlation. The intrinsic correlation is found to resemble a linear relation well, with a slope of $\sim$6, independent of the chosen $R$ and $\Gamma$. When we tried to fit the data with a line, we found that the best-fit slope is different from that of the intrinsic correlation at a level $>99.9\%$. We therefore conclude that the observed correlation is genuine. The reality of the $R-\Gamma$ correlation has been discussed in detail and verified by \cite{2003MNRAS.342..355Z}.

This correlation, firstly observed by \cite{1999MNRAS.303L..11Z}, has been explained to be a result of the relative motion of the primary X-ray source with respect to the disc, which is now assumed to be the reflector. When the X-ray source moves away from the disc with a slightly relativistic velocity, the emission is beamed away from the disc, resulting in fewer photons to be reflected, and vice versa when the corona approaches the disc. \cite{1999ApJ...510L.123B} showed that the Compton amplification factor, A, and the reflection strength can be described as functions of the corona velocity and of a geometry-dependent parameter, called $\mu_s$. He also estimated that the spectral index is related to A as $\Gamma \simeq 2.33 (A-1)^{-\delta}$ with $\delta \simeq 0.1$ for the AGN. Using these equations and for an albedo $\alpha = 0.15$, a system inclination of 30 degrees, and $\mu_s=0.55$, we calculated the expected curve, which is plotted as the dashed blue line in Fig. \ref{fig:gamma_vs_refl}. This line represents the observations well although it was not fitted to them. It should be noted that the predicted reflection has been estimated by multiplying Eq. (3) of the aforementioned paper with the cosine of the inclination angle $\mu = cos\theta$ to account for the viewing angle that affects the reflected emission but not the coronal emission (this is a small correction for face-on sources).

An alternative explanation for the observed correlation can be given if the disc-covering factor, $f_c$, towards the corona is let free to vary. This is expected if the disc is patchy or if its inner radius varies. \cite{2018A&A...614A..79P} (see their Fig. 1 for the various geometries) have calculated the power-law slope expected for different covering factors (Fig. 5 in their paper). In addition, the covering factor for a specific inclination is directly proportional to the reflection strength. Using the results of \cite{2018A&A...614A..79P} and assuming $R=2\cdot f_{c}$ \citep[to take into account anisotropic inverse Compton emission,][]{1991MNRAS.248...14G}, we calculated the expected $R-\Gamma$ curve, plotted as a red dot-dashed line in Fig. \ref{fig:gamma_vs_refl}. The estimated curve resembles the observed trend. Disc dissipation allows us to reach $\Gamma>2.2$. A more detailed analysis is required that would self-consistently calculate $\Gamma$ and $R$ for a range of covering factors and inclinations.

Regardless of the chosen interpretation, the $R-\Gamma$ correlation strongly favours a disc origin for the Compton hump in unabsorbed sources. \cite{2013MNRAS.431.2441D} and \cite{2016MNRAS.462..511K} explored the XMM light curves of several AGN for soft (i.e. 0.3-1.0 keV) or Fe K$\alpha$ lags, respectively. We cross-matched our sample with the sources of these works and found that 13 of our sources have been shown to exhibit a soft or Fe lag, or both; 12 of them were categorised into the first two classes. The soft and Fe lags are produced when the primary X-ray continuum is reflected by the disc, and thus, their detection further supports the assumption that the disc is the source of Compton hump for Classes 1 and 2.

The question that arises from this discussion is whether the torus contributes to the observed reflection as well. We show in the next section that the origin of reflection for obscured sources is the torus. However, the obscured and unobscured sources have a similar range of reflection strength (see Fig. \ref{fig:best_fit}). Therefore, the contribution of torus reflection in Classes 1 and 2 should be negligible, if any, suggesting an intrinsic difference between the absorbed and unabsorbed sources. This result could be explained if the torus in unobscured sources was thinner or if it covered a smaller area of the sky as seen by the corona.

\subsection{Absorbed sources (Classes 3 and 4)}

\begin{figure}
  \centering
  \includegraphics[width=\linewidth,clip]{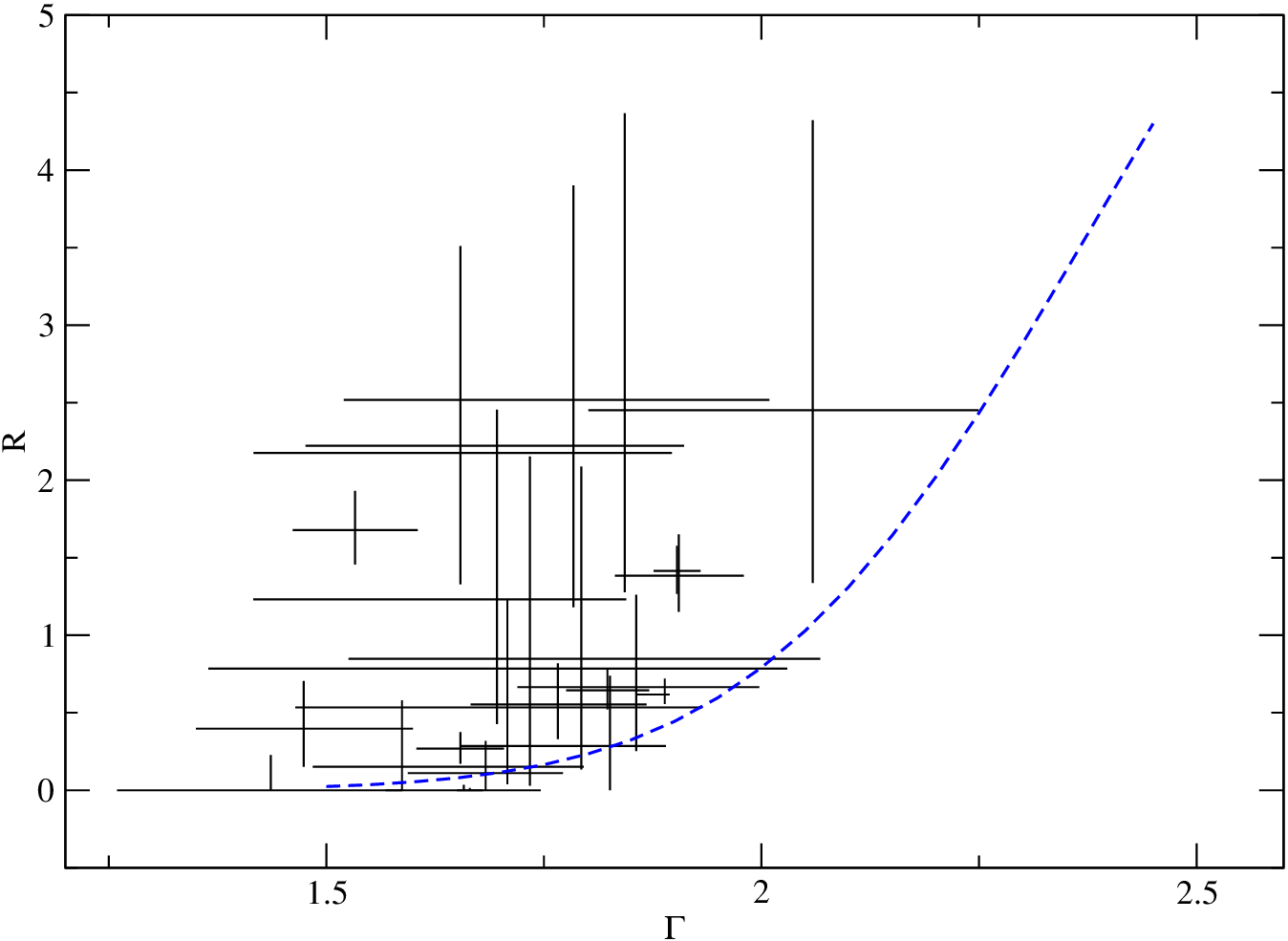}
  \caption[9]{Best-fit reflection vs. power-law slope for the objects in Classes 3 and 4. The blue dashed line indicates the theoretical curve presented in Fig. \ref{fig:gamma_vs_refl}.}
  \label{fig:gamma_vs_refl_class34}
\end{figure}

Figure \ref{fig:gamma_vs_refl_class34} plots the best-fit reflection values as a function of the power-law slope for Classes 3 and 4. The blue dashed line denotes the same theoretical curve as we plotted in Fig. \ref{fig:gamma_vs_refl}. Despite the existence of a group of sources with high reflection, there is no evidence for a correlation between $R$ and $\Gamma$ and the values do not follow the curve that is observed for the unabsorbed sources. As a result, the reflection increase in these objects cannot be explained by the motion of the corona. We investigated whether the reflection variations could be associated with the obscuration of each object.

\begin{figure}
  \centering
  \includegraphics[width=\linewidth,clip]{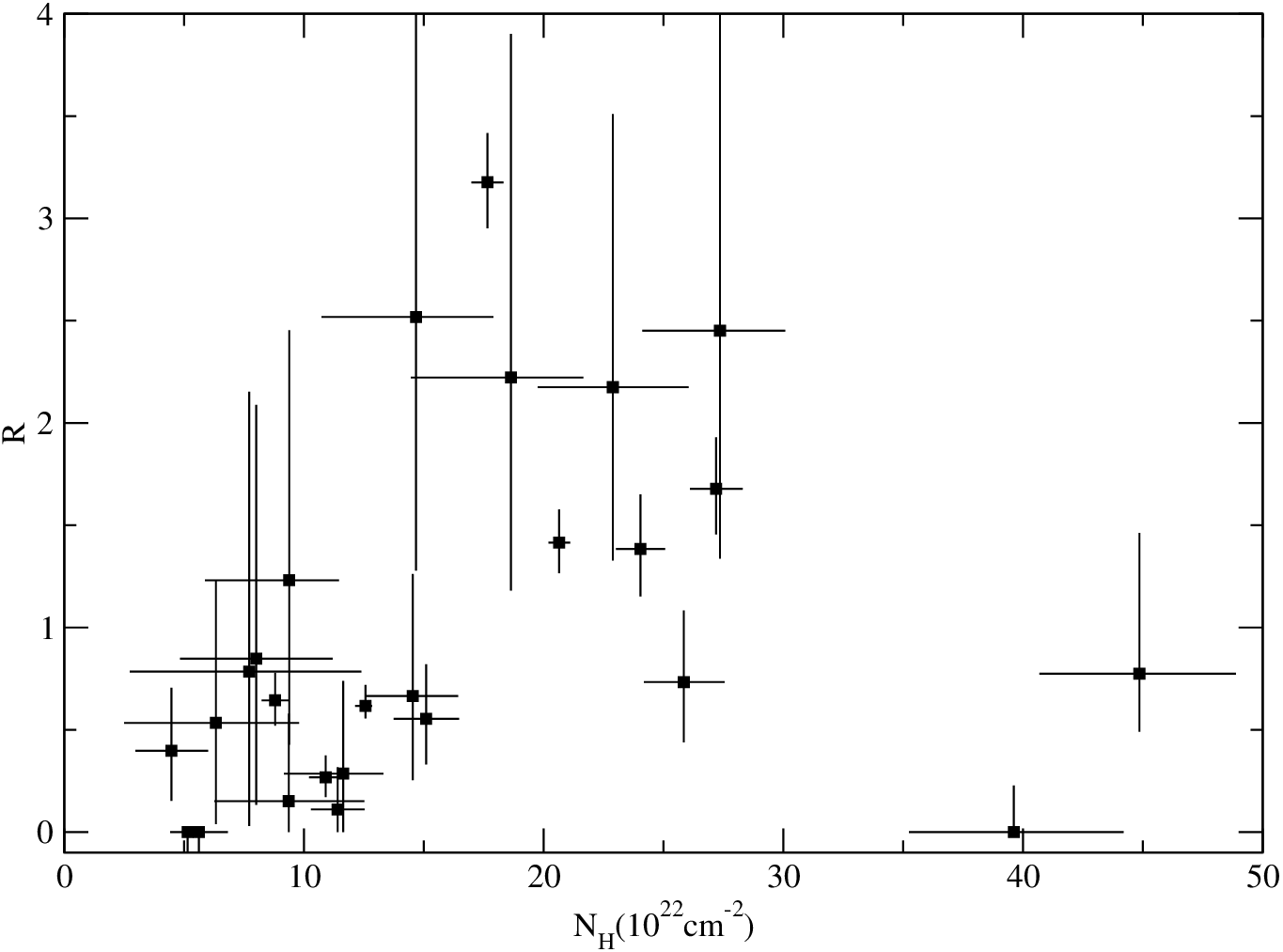}
  \caption[9]{Correlation between reflection and absorption in the sources of Classes 3 and 4.}
  \label{fig:refl_vs_nh}
\end{figure}

Figure \ref{fig:refl_vs_nh} plots the best-fit reflection versus absorption for Classes 3 and 4. The reflection seems to be at least slightly correlated to the absorption column density for densities up to $3 \cdot 10^{23} $cm$^{-2}$. This result is expected because it has already been observed for the average properties of the two classes we considered here. A tentative indication of a positive $R-N_H$ has also been found by \cite{2017ApJ...849...57D}.

The lack of a strong $R-\Gamma$ correlation and the observed dependence of $R$ on $N_H$ are not easily justified if the disc is the main reflector in these sources. Instead, most of the observed reflection should take place in the putative torus.

The observed $R-N_H$ correlation is difficult to explain when a simple geometry is assumed for the torus. The estimated $R$ values exclude the possibility of a Compton-thin torus, that is, a torus for which the equatorial column density, $N_{H,eq}$, is lower than $1.5 \cdot 10^{24} cm^{-2}$; because then $R$ should be significantly smaller than 1.

In the case of a CT torus with a common $N_{H,eq}$ for all the sources, the reflection would be expected to be anti-correlated to the line-of-sight absorption, $N_{H}$. When a source is observed at a higher inclination angle, the observed column density is higher and the strength of reflection should be lower because a smaller area of the torus is visible to us. Because we found a positive correlation between $R$ and $N_H$, this possibility is also excluded.

Finally, we explored whether such a correlation is predicted when we consider different equatorial absorption levels for each source. A denser torus will result in a higher $R$ value because the continuum emission is more strongly suppressed. While tempting, this model becomes less plausible when we take the inclination angle of the system into account. Using the equation

\begin{eqnarray}
  \centering
      N_{H} = N_{H,eq} \left[ 1 - {\left(\frac{c}{a}\right)}^2 \cos^2\theta_{obs} \right]^{1/2}  
,\end{eqnarray}

\noindent where $c$ and $a$ indicate the distance of the torus from the central black hole and the radius of the torus, respectively, we estimated the inclination angle expected for the sources of Fig. \ref{fig:refl_vs_nh}. We assumed $\frac{c}{a} = 2$ and $N_{H,eq} > 1.5 \cdot 10^{24} cm^{-2}$. Thus, we found that all the sources need to be observed through the edge of the torus with an inclination angle of 60 to 61 degrees, which seems highly unlikely.

A clumpy torus with a population of small clouds rotating around the black hole in a toroidal distribution provides a more natural explanation for the $R-N_H$ relation. The latter could be associated with the cloud filling factor. A source with a larger distribution of clouds will be more obscured because more clouds lie on the line of sight. At the same time, the reflection strength will be higher as well because the scattering surface increases.

We have temporarily excluded from this discussion the two sources with $N_H > 35 \cdot 10^{22} cm^{-2}$ of Fig. \ref{fig:refl_vs_nh}. These two outliers exhibit high absorption levels and little or no reflection. In the case of a clumpy torus, when the number of clouds increases significantly, the clouds fill most of the surrounding space in a roughly continuous way, and their effect can be approximated as a compact torus. Then we expect the absorption to be high and the reflection to drop to lower values, as observed for these two objects.

The obscured sources have on average a smaller spectral index than the unabsorbed sources (Fig. \ref{fig:best_fit}).  $\Gamma$ ranges from $\sim 1.5$ to 2.5 for Classes 1 and 2 and from $\sim 1.4$ to 2.0 for Classes 3 and 4. While this might indicate intrinsic differences in the corona of the two samples, it can be explained by geometrical differences. The higher slopes suggest a lower optical depth for the corona of unabsorbed sources. If the corona was homogeneous and similar in all the sources, a lower optical depth would be obtained when the photons travel through a smaller distance within the corona. When a higher inclination angle is assumed for the absorbed sources, the differences in the observed $\Gamma$ can be explained if the corona has a geometry close to a slab. In this case, photons observed from sources viewed close to edge-on travel through a thicker corona than those observed in sources that are observed face-on, and thus correspond to a smaller $\Gamma$.

We suggested above that the main reflector in absorbed sources is a clumpy torus, which gives rise to the question why the accretion disc does not contribute much to the observed reflection in these sources. If obscured sources are observed at higher inclination angles on average, it is expected that the disc reflection appears fainter with respect to the coronal emission.

\subsection{Compton-thick sources}

The more strongly obscured sources were found to be in Class 5. Most of them have a reflection-dominated spectrum and a best-fit $N_H$ consistent with a CT medium. Our spectral analysis suggested that these objects can be further divided into two groups.

We found that a simple torus model can fit only 60 \% of these sources well, while the rest require a more complicated model. This indicates a different reflector geometry. The dusty region might be more extended or not circular. It is also possible that these sources feature an additional reflection of the primary continuum from an absorbing region outside the nucleus. This region could be part of the host galaxy located at a distance of several tens or hundreds of parsec from the black hole, as has been imaged by Chandra in a few obscured AGN \citep[e.g.][]{2015ApJ...812..116B,2018ApJ...865...83F}.

\subsection{Fe K$\alpha$ line}

A detailed study of the Fe line in the various sources is beyond the scope of this work. Nevertheless, we used a simple Gaussian model to simulate the observed emission in the spectral fitting. We were able to calculate the line equivalent width (EW) for each source. The mean EW values for each class are listed in Table \ref{tab:eqws}.

\begin{table}
        \centering
        \caption{Mean values of the Fe K$\alpha$ EW in each class.}
        \label{tab:eqws}
        \begin{tabular}{lc} 
                \hline
                          & Fe K$\alpha$ <EW> (keV) \\
                \hline
                Class 1   & 0.125 $\pm$  0.025    \\
                Class 2   & 0.098 $\pm$  0.011    \\
                Class 3   & 0.074 $\pm$  0.010    \\
                Class 4   & 0.119 $\pm$  0.024    \\
                \hline  
        \end{tabular}
        \vspace{1cm}
\end{table}

\begin{figure}
  \centering
  \includegraphics[width=\linewidth,clip]{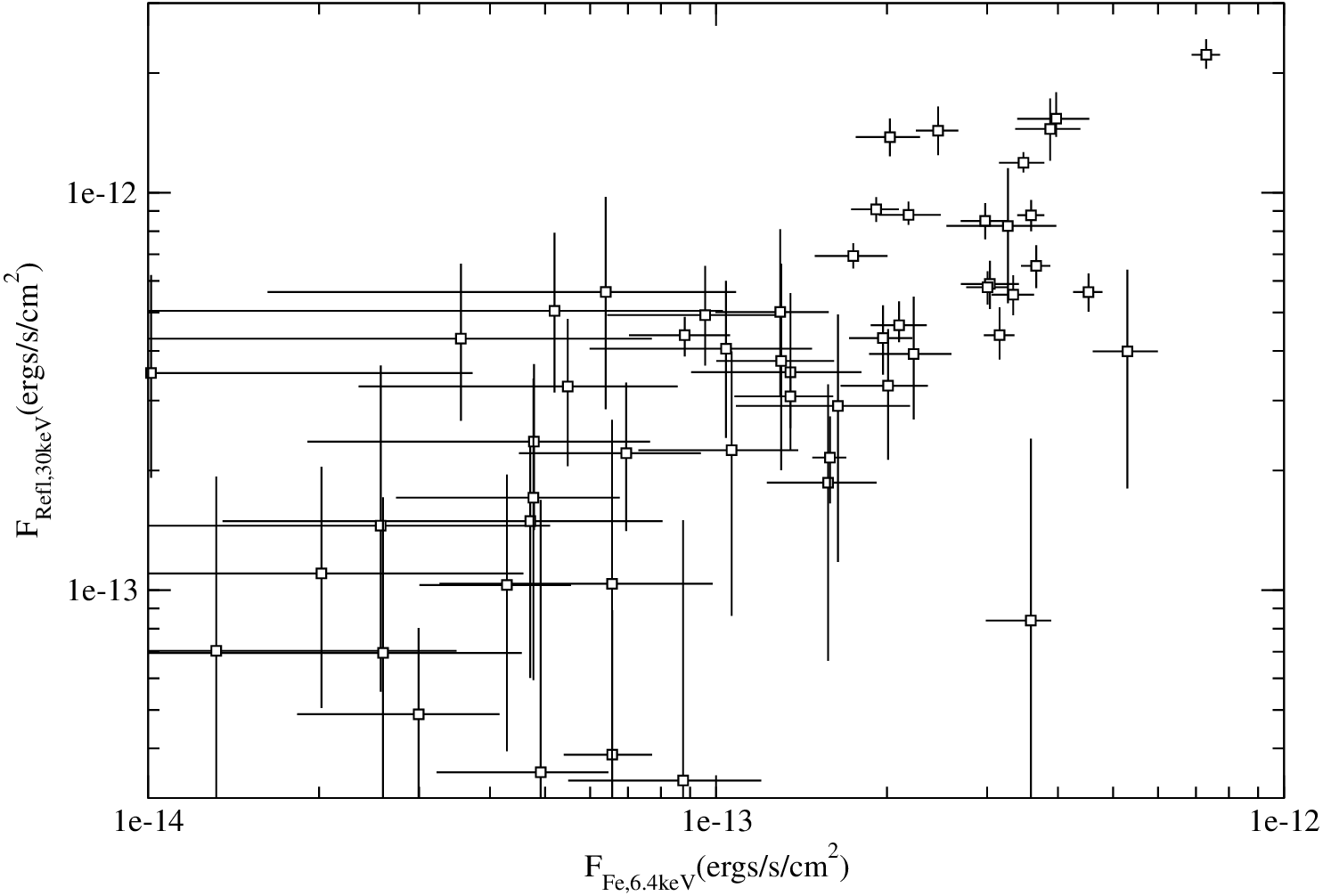}
  \caption[9]{Flux of the reflected emission vs. the flux of the Fe K$\alpha$ line in Classes 1-4, as estimated using the best-fit results. Both axes are in logarithmic scale.}
  \label{fig:reflux_vs_feflux}
\end{figure}

The average Fe line EW seems to be similar in all the different classes, but exhibits a large variability in each class. This result indicates that the Fe emission is not associated with the absorption or the power-law slope.

Because the Compton hump and the Fe line are both thought to be produced by the reflection of the primary continuum, we would expect the two emissions to be positively correlated. For each source, we used the best-fit parameters to estimate the Fe line flux at 6.4 keV (in the AGN rest frame) and the flux of the reflected emission at 30 keV.

The calculated fluxes are plotted in Fig. \ref{fig:reflux_vs_feflux}. Their clear correlation shows that the Fe line and the Compton hump are produced by the scattering of the primary continuum by the same material.

\section{Summary}

We studied the NuSTAR spectra of a sample of local AGN. Based on their X-ray spectral shape, we classified the sources into several categories and fitted their SED with an appropriate model.

We confirmed, using both individual and stacked spectra, that mildly obscured sources ($23<log(N_H)<24$) exhibit a great reflection strength, with an average value of $\langle R \rangle =1.7 \pm 0.3$. This result constrains the fraction of CT AGN in the universe and challenges the unification model. The reflection strength of the lightly absorbed sources (Class 3) was found to be $\langle R \rangle =0.5 \pm 0.1$.
  
The reflection strength is correlated to the photon index in unabsorbed sources. This correlation can be naturally explained by the relative motion of the corona with respect to the disc or by a variation in disc covering factor. Thus, the disc is the main reflector in these sources. This result  is further supported by the detection of soft and Fe lags in some of these sources.

For the absorbed sources, we found evidence of a correlation between the reflection and the absorption column density, suggesting that the reflection takes place at larger distances in these objects. The observed correlation cannot be easily understood, unless a clumpy torus is assumed. The cloud filling factor probably drives the correlation, and a more detailed model is required to reproduce the observations.

Almost half of the CT AGN (8 out of 20) require an absorbing region that is more complex than a compact homogeneous torus to reproduce the observed spectra well. It is possible that reflection occurs in a larger region in these objects.
 
The Fe K$\alpha$ line is positively correlated to the flux of Compton hump. This indicates that the two emissions are produced by the same medium.

\begin{table*}
        \centering
        \caption{The used source sample.}
        \label{tab:sources_info}
        \begin{tabular}{lcccccccrr} 
                \hline
                {Source Name}                &  RA                       &Dec             &z      &Type   &Class     &Obs ID        &Obs date      &Exposure     & source id \\
                                             &                           &                &       &       &          &              & (MJD)        &(ks)                     \\
                \hline
                1H 0323+342                   & 51.1715                   & 34.1794      & 0.0610  & NLS1     & 1       & 60061360002  &  56731.742      & 100.95   & 1   \\
    4U 0937-12                    & 146.4252                  & -14.3264     & 0.0077  & Seyf 1   & 1       & 60160371002  &  57358.578      &  20.80   & 2   \\
                IGR J12415-5750               & 190.3573                  & -57.8343     & 0.0244  & Seyf 1.5 & 1       & 60160510002  &  57505.853      &  16.37   & 3   \\
                IGR J16482-3036               & 252.0635                  & -30.5845     & 0.0310  & Seyf 1   & 1       & 60160648002  &  57846.308      &  17.48   & 4   \\
                IGR J18027-1455               & 270.6896                  & -14.9089     & 0.0350  & Seyf 1   & 1       & 60160680002  &  57509.424      &  19.88   & 5   \\
                IGR J19077-3925               & 286.9600                  & -39.3921     & 0.0725  & Seyf 2   & 1       & 60061291002  &  57535.543      &  16.34   & 6   \\
                1RXS J211928.4+333259         & 319.8675                  & 33.5514      & 0.0510  & Seyf 1.5 & 1       & 60061358002  &  57039.697      &  19.30   & 7   \\
                Mrk 1018                      & 31.5666                   & -0.2914      & 0.0424  & Seyf 2   & 1       & 60160087002  &  57428.939      &  21.62   & 8   \\
          Mrk 520                       & 330.1746                  & 10.5497      & 0.0266  & Seyf 2   & 1       & 60160774002  &  57883.112      &  20.90   & 9   \\
                Mrk 590                       & 33.6398                   & -0.7666      & 0.0264  & Seyf 1   & 1       & 60160095002  &  57423.702      &  21.21   & 10  \\
                                              &                           &              &         &          &         & 90201043002  &  57724.495      &  50.62         \\
                PG 1501+106                   & 226.0050                  & 10.4378      & 0.0364  & Seyf 1.5 & 1       & 60101023002  &  57217.742      &  17.70   & 11  \\
                
                Ark 120                       & 79.0476                   & -0.1498      & 0.0327  & Seyf 1   & 2       & 60001044002  &  56341.465      &  55.24   & 12  \\
                                              &                           &              &         &          &         & 60001044004  &  56738.410      &  64.70 \\
                ESO 141-55                    & 290.3089                  & -58.6703     & 0.0371  & Seyf 1   & 2       & 60201042002  &  57584.694      &  92.09   & 13  \\
                ESO 33-2                      & 73.9957                   & -75.5412     & 0.0181  & Seyf 2   & 2       & 60061054002  &  56781.032      &  23.04   & 14   \\
                Fairall 1146                  & 129.6283                  & -35.9926     & 0.0316  & Seyf 1   & 2       & 60061082002  &  56865.966      &  21.28   & 15   \\
                1H 2251-179                   & 343.5242                  & -17.5819     & 0.0640  & Seyf 1.5 & 2       & 60102025002  &  57160.188      &  22.95   & 16   \\
                                              &                           &              &         &          &         & 60102025004  &  57190.151      &  22.86      \\
                                              &                           &              &         &          &         & 60102025006  &  57336.992      &  19.45      \\
                                              &                           &              &         &          &         & 60102025008  &  57367.684      &  20.68      \\                                 
          IGR J14471-6414               & 221.6164                  & -64.2731     & 0.0530  & Seyf 1   & 2       & 60061257002  &  56440.318      &  15.04   & 17   \\
                IGR J14552-5133               & 223.8242                  & -51.5714     & 0.0160  & NLS1     & 2       & 60061259002  &  56554.660      &  21.90   & 18  \\
                SWIFT J1650.5+0434            & 252.6781                  &  4.6050      & 0.0320  & Seyf 2   & 2       & 60061273002  &  57790.725      &  21.03   & 19  \\
                SWIFT J1933.9+3258            & 293.4465                  & 32.9072      & 0.0565  & Seyf 1   & 2       & 60160714002  &  57539.642      &  12.65   & 20  \\
                IGR J19378-0617               & 294.3875                  & -6.2180      & 0.0102  & NLS1     & 2       & 60101003002  &  57296.750      &  65.02   & 21  \\
                IGR J21277+5656               & 321.9372                  & 56.9443      & 0.0144  & NLS1     & 2       & 60001110002  &  56235.737      &  49.06   & 22  \\
                                              &                           &              &         &          &         & 60001110003  &  56236.755      &  28.49     \\
                                              &                           &              &         &          &         & 60001110005  &  56237.758      &  62.43      \\
                                              &                           &              &         &          &         & 60001110007  &  56239.711      &  42.11   \\                                                           
                MCG -06-30-015                & 203.9738                  & -34.2955     & 0.0078  & Seyf 1.5 & 2       & 60001047002  &  56321.478      &  23.27   & 23  \\
                                              &                           &              &         &          &         & 60001047003  &  56322.016      & 125.86   \\
                                              &                           &              &         &          &         & 60001047005  &  56325.452      &  29.49   \\                                      
                Mrk 1040                      & 37.0603                   & 31.3117      & 0.0167  & Seyf 1   & 2       & 60101002002  &  57246.997      &  50.03   & 24   \\
                                              &                           &              &         &          &         & 60101002004  &  57249.208      &  49.75   \\
                Mrk 509                       & 311.0406                  & -10.7235     & 0.0344  & Seyf 1.5 & 2       & 60101043002  &  57141.602      & 165.50   & 25   \\
                                              &                           &              &         &          &         & 60101043004  &  57175.433      &  36.48   \\
                Mrk 766                       & 184.6105                  & 29.8129      & 0.0129  & NLS1     & 2       & 60001048002  &  57046.525      &  89.36   & 26   \\
                                              &                           &              &         &          &         & 60101022002  &  57208.740      &  16.30   \\
                NGC 0985                      & 38.6574                   & -8.7876      & 0.0431  & Seyf 1.5 & 2       & 60061025002  &  56515.496      &  13.89   & 27   \\
                NGC 3516                      & 166.6979                  & 72.5686      & 0.0088  & Seyf 1.5 & 2       & 60002042002  &  56832.702      &  50.42   & 28   \\
                                              &                           &              &         &          &         & 60002042004  &  56849.634      &  71.33   \\
                NGC 4051                      & 180.7901                  & 44.5313      & 0.0023  & NLS1     & 2       & 60001050002  &  56460.706      &   9.43   & 29   \\
                                              &                           &              &         &          &         & 60001050003  &  56460.908      &  45.74   \\
                                              &                           &              &         &          &         & 60001050005  &  56574.576      &   9.39   \\
                                              &                           &              &         &          &         & 60001050006  &  56574.846      &  48.95   \\
                                              &                           &              &         &          &         & 60001050008  &  56704.574      &  56.55   \\
                NGC 4593                      & 189.9143                  & -5.3443      & 0.0090  & Seyf 1   & 2       & 60001149002  &  57020.042      &  22.10   & 30   \\
                                              &                           &              &         &          &         & 60001149004  &  57022.193      &  20.79   \\
                                              &                           &              &         &          &         & 60001149006  &  57024.143      &  20.56   \\
                                              &                           &              &         &          &         & 60001149008  &  57026.092      &  22.10   \\
                                              &                           &              &         &          &         & 60001149010  &  57028.646      &  21.14   \\
                NGC 5506                      & 213.3121                  & -3.2076      & 0.0062  & NLS1     & 2       & 60061323002  &  56748.996      &  56.59   & 31   \\
                NGC 5548                      & 214.4981                  & 25.1368      & 0.0172  & Seyf 1.5 & 2       & 60002044002  &  56484.419      &  22.77   & 32   \\
                                              &                           &              &         &          &         & 60002044003  &  56485.026      &  26.96   \\
                                              &                           &              &         &          &         & 60002044005  &  56496.613      &  49.39   \\
                                              &                           &              &         &          &         & 60002044006  &  56545.904      &  51.11   \\
                                              &                           &              &         &          &         & 60002044008  &  56646.357      &  50.10   \\
                
                \hline  
        \end{tabular}
\end{table*}            
                
\addtocounter{table}{-1}                
                
\begin{table*}
        \centering
        \caption{Continued.}
        \label{tab:sources_info}
        \begin{tabular}{lcccccccrr} 
                \hline
                {Source Name}                &  RA                       &Dec             &z      &Type   &Class     &Obs ID        &Obs date      &Exposure     & source id     \\
                                             &                           &                &       &       &          &              & (MJD)        &(ks)        \\
                \hline                          
                NGC 5995                      & 237.1040                  & -13.7578     & 0.0252  & Seyf 2   & 2       & 60061267002  &  56897.201      &  21.18   & 33   \\
                NGC 6814                      & 295.6694                  & -10.3235     & 0.0052  & Seyf 1.5 & 2       & 60201028002  &  57573.751      & 143.12   & 34   \\
                NGC 7314                      & 338.9425                  & -26.0505     & 0.0048  & Seyf 2   & 2       & 60201031002  &  57521.521      &  99.01   & 35   \\
                NGC 7469                      & 345.8151                  &  8.8740      & 0.0163  & Seyf 1.5 & 2       & 60101001002  &  57185.785      &  21.58   & 36   \\
                                              &                           &              &         &          &         & 60101001004  &  57350.693      &  20.03   \\
                                              &                           &              &         &          &         & 60101001006  &  57371.446      &  16.42   \\
                                              &                           &              &         &          &         & 60101001008  &  57378.430      &  18.98   \\
                                              &                           &              &         &          &         & 60101001010  &  57381.589      &  20.68   \\
                                              &                           &              &         &          &         & 60101001012  &  57383.064      &  19.23   \\
                                              &                           &              &         &          &         & 60101001014  &  57384.945      &  23.40   \\
                PG 0804+761                   & 122.7442                  &  76.0451     & 0.1000  & Seyf 1   & 2       & 60160322002  &  57480.117      &  16.94   & 37   \\
                
                SWIFT J0318.7+6828            &  49.5791                  &  68.4921     & 0.0901  & Seyf 2   & 3       & 60061342002  &  57512.802      &  24.04   & 38   \\
                SWIFT J0505.7-2348            &  76.4405                  & -23.8539     & 0.0350  & Seyf 2   & 3       & 60061056002  &  56525.403      &  21.16   & 39   \\
                SWIFT J0920.8-0805            & 140.1927                  &  -8.0561     & 0.0196  & Seyf 2   & 3       & 60061091002  &  56385.532      &  12.38   & 40   \\
                                              &                           &              &         &          &         & 60061091004  &  56392.605      &   9.39   \\
                                              &                           &              &         &          &         & 60061091006  &  56400.693      &  11.59   \\
                                              &                           &              &         &          &         & 60061091008  &  56417.729      &  14.09   \\
                                              &                           &              &         &          &         & 60061091010  &  56424.532      &  15.33   \\
                                              &                           &              &         &          &         & 60061091012  &  56434.298      &  12.29   \\
                IGR J09253+6929               & 141.4481                  &  69.4648     & 0.0398  & Seyf 1.5 & 3       & 60201030002  &  57569.884      &  44.10   & 41   \\
                IGR J12391-1612               & 189.7762                  & -16.1797     & 0.0367  & Seyf 2   & 3       & 60061232002  &  57402.305      &  21.35   & 42   \\
                IGR J19473+4452               & 296.8307                  &  44.8284     & 0.0539  & Seyf 2   & 3       & 60061292002  &  56237.349      &  18.21   & 43   \\
                IGR J20216+4359               & 305.4544                  &  44.0110     & 0.0170  & Seyf 2   & 3       & 60061298002  &  56933.670      &  21.10   & 44   \\
                IGR J23308+7120               & 352.6570                  &  71.3796     & 0.0370  & Seyf 2   & 3       & 60061320002  &  56956.670      &  16.08   & 45   \\
                IGR J23524+5842               & 358.0922                  &  58.7586     & 0.1640  & Seyf 2   & 3       & 60160838002  &  57384.476      &  22.38   & 46   \\
                Mrk 348                       &  12.1964                  &  31.9570     & 0.0150  & Seyf 2   & 3       & 60160026002  &  57323.296      &  21.52   & 47   \\
                NGC 2110                      &  88.0474                  &  -7.4562     & 0.0078  & Seyf 2   & 3       & 60061061002  &  56205.233      &  15.54   & 48   \\
                                              &                           &              &         &          &         & 60061061004  &  56337.088      &  11.51   \\
                NGC 4258                      & 184.7396                  &  47.3040     & 0.0015  & Seyf 2   & 3       & 60101046002  &  57342.466      &  53.00   & 49   \\
                                              &                           &              &         &          &         & 60101046004  &  57397.133      & 102.08   \\
                NGC 4395                      & 186.4536                  &  33.5469     & 0.0011  & Seyf 2   & 3       & 60061322002  &  56422.114      &  18.95   & 50   \\
                NGC 5252                      & 204.5665                  &   4.5426     & 0.0230  & Seyf 2   & 3       & 60061245002  &  56423.072      &  19.01   & 51   \\
                NGC 7172                      & 330.5079                  & -31.8697     & 0.0087  & Seyf 2   & 3       & 60061308002  &  56937.572      &  31.67   & 52   \\
                
                ESO 103-35                    & 279.5848                  & -65.4276     & 0.0133  & Seyf 2   & 4       & 60061288002  &  56347.893      &  27.39   & 53   \\
                IGR J01528-0326               &  28.2042                  &  -3.4468     & 0.0172  & Seyf 2   & 4       & 60061016002  &  56261.015      &  13.42   & 54   \\
                IGR J06239-6052               &  95.9399                  & -60.9790     & 0.0405  & Seyf 2   & 4       & 60061065002  &  56877.124      &  20.21   & 55   \\
                SWIFT J1049.4+2258            & 162.3789                  &  22.9646     & 0.0328  & Seyf 2   & 4       & 60061206002  &  57804.539      &  20.69   & 56   \\
                IGR J14579-4308               & 224.4288                  & -43.1317     & 0.0157  & Seyf 2   & 4       & 60061260002  &  56506.510      &   7.79   & 57   \\
                IGR J18244-5622               & 276.0808                  & -56.3692     & 0.0169  & Seyf 2   & 4       & 60061284002  &  56497.778      &  19.83   & 58   \\
                SWIFT J1930.5+3414            & 292.5575                  &  34.1804     & 0.0616  & Seyf 1.5 & 4       & 60160713002  &  57588.121      &  20.47   & 59   \\
                Mrk 477                       & 220.1587                  &  53.5044     & 0.0377  & Seyf 2   & 4       & 60061255002  &  56792.194      &  18.08   & 60   \\
                                              &                           &              &         &          &         & 60061255004  &  56801.208      &  16.80   \\
                Mrk 6                         & 103.0510                  &  74.4271     & 0.0188  & Seyf 1.5 & 4       & 60102044002  &  57133.119      &  62.48   & 61   \\
                                        &                           &              &         &          &         & 60102044004  &  57335.123      &  43.18   \\
                NGC 6300                      & 259.2478                  & -62.8206     & 0.0037  & Seyf 2   & 4       & 60061277002  &  56348.898      &  17.04   & 62   \\
                                              &                           &              &         &          &         & 60261001002  &  57411.047      &  20.05   \\
                                              &                           &              &         &          &         & 60261001004  &  57624.369      &  22.29   \\
                NGC 7582                      & 349.5979                  & -42.3706     & 0.0053  & CT       & 4       & 60061318002  &  56170.712      &  16.42   & 63   \\
                                              &                           &              &         &          &         & 60061318004  &  56184.730      &  14.10   \\
                                              &                           &              &         &          &         & 60201003002  &  57506.215      &  47.07   \\
                
                IGR J00254+6822               &   6.3870                  &  68.3622     & 0.0120  & Seyf 2   & 5       & 60061003002  &  56758.161      &  26.01   & 64   \\
                IGR J02501+5440               &  42.6775                  &  54.7049     & 0.0150  & Seyf 2   & 5       & 60061030002  &  56339.637      &  15.56   & 65   \\
                SWIFT J0453.4+0404            &  73.3573                  &   4.0616     & 0.0294  & Seyf 2   & 5       & 60061053002  &  56318.963      &  10.00   & 66   \\
                                              &                           &              &         &          &         & 60061053004  &  56882.774      &  18.30   \\
                                              &                           &              &         &          &         & 60001158002  &  56971.279      &  94.88   \\
                SWIFT J0601.9-8636            &  91.4235                  & -86.6319     & 0.0062  & CT       & 5       & 60061063002  &  57336.553      &  23.89   & 67   \\
                SWIFT J1009.3-4250            & 152.4509                  & -42.8112     & 0.0335  & Seyf 2   & 5       & 60061098002  &  57308.540      &  18.46   & 68   \\
                IGR J13091+1137               & 197.2734                  &  11.6342     & 0.0251  & CT       & 5       & 60061239002  &  57049.216      &  23.36   & 69   \\
                IGR J14175-4641               & 214.2653                  & -46.6948     & 0.0766  & Seyf 2   & 5       & 60201033002  &  57533.385      &  21.24   & 70   \\
                IGR J16351-5806               & 248.8088                  & -58.0800     & 0.0091  & CT       & 5       & 60061272002  &  57546.887      &  18.55   & 71   \\
                IGR J20286+2544               & 307.1461                  &  25.7333     & 0.0139  & Seyf 2   & 5       & 60061300002  &  56430.285      &  19.23   & 72   \\
                
                \hline  
        \end{tabular}
\end{table*}

\addtocounter{table}{-1}                
                
\begin{table*}
        \centering
        \caption{Continued.}
        \label{tab:sources_info}
        \begin{tabular}{lcccccccrr} 
                \hline
                {Source Name}                &  RA                       &Dec             &z      &Type   &Class     &Obs ID        &Obs date      &Exposure     & source id        \\
                                             &                           &                &       &       &          &              & (MJD)        &(ks)        \\
                \hline                          
                LEDA 15023                    &  65.9199                  &   4.1338     & 0.0450  & Seyf 2   & 5       & 60006005001  &  56132.851      &   6.24   & 73   \\
                                              &                           &              &         &          &         & 60006005002  &  56133.055      &   6.02   \\
                                              &                           &              &         &          &         & 60006005003  &  56133.458      &   5.44   \\
                LEDA 96373                    & 111.6098                  & -35.9060     & 0.0294  & Seyf 2   & 5       & 60061073002  &  56869.865      &  22.02   & 74   \\
                MCG -03-34-064                & 200.6019                  & -16.7285     & 0.0165  & Seyf 2   & 5       & 60101020002  &  57404.322      &  77.96   & 75   \\
                Mrk 3                         &  93.9015                  &  71.0375     & 0.0135  & CT       & 5       & 60002048002  &  56907.727      &  30.02   & 76   \\
                                              &                           &              &         &          &         & 60002048004  &  56914.452      &  33.48    \\
                                              &                           &              &         &          &         & 60002048006  &  56931.529      &  33.18    \\
                                              &                           &              &         &          &         & 60002048008  &  56939.196      &  25.79    \\
                                              &                           &              &         &          &         & 60002048010  &  56953.253      &  30.82    \\
                                              &                           &              &         &          &         & 60002048012  &  57104.243      &  26.15    \\
                                              &                           &              &         &          &         & 60002049002  &  57100.547      &  21.70    \\
                                              &                           &              &         &          &         & 60002049004  &  57117.076      &  24.51    \\
                                              &                           &              &         &          &         & 60002049006  &  57120.368      &  25.11    \\
                                              &                           &              &         &          &         & 60002049008  &  57125.540      &  24.87    \\
                                              &                           &              &         &          &         & 60002049010  &  57132.650      &  27.23    \\

                NGC 1194                      &  45.9546                  &  -1.1037     & 0.0136  & Seyf 2   & 5       & 60061035002  &  57081.456      &  26.09   & 77   \\
                NGC 3281                      & 157.9670                  & -34.8537     & 0.0107  & CT       & 5       & 60061201002  &  57409.955      &  22.99   & 78   \\
                NGC 4507                      & 188.9026                  & -39.9093     & 0.0118  & Seyf 2   & 5       & 60102051002  &  57145.851      &  30.13   & 79  \\
                                              &                           &              &         &          &         & 60102051004  &  57183.810      &  33.65  \\
                                              &                           &              &         &          &         & 60102051006  &  57218.341      &  30.39   \\
                                              &                           &              &         &          &         & 60102051008  &  57256.427      &  30.63   \\

                NGC 4941                      & 196.0548                  &  -5.5516     & 0.0037  & Seyf 2   & 5       & 60061236002  &  57406.078      &  15.78   & 80   \\
                NGC 5643                      & 218.1698                  & -44.1744     & 0.0040  & CT       & 5       & 60061362002  &  56801.611      &  22.46   & 81   \\
                                              &                           &              &         &          &         & 60061362004  &  56838.474      &  19.33   \\
                NGC 5728                      & 220.5996                  & -17.2531     & 0.0094  & CT       & 5       & 60061256002  &  56294.203      &  24.10   & 82   \\
                NGC 788                       &  30.2769                  &  -6.8155     & 0.0136  & Seyf 2   & 5       & 60061018002  &  56320.235      &  15.41   & 83   \\
                
                \hline  
        \end{tabular}
        \tablefoot{The first column lists the source name as given in R11; the second and third columns denote the celestial coordinates, and the fourth column lists the redshift of the AGN. The fifth and sixth column list the traditional classification of the sources and the classification used here, respectively. Finally, Columns 7 to 9 provide information of the NuSTAR observations we analysed here, and the last column gives the source identification number we used.}
\end{table*}

\begin{table*}
        \centering
        \caption{Best-fit parameters for all the sources in Classes 1 to 4.}
        \label{tab:best_fit}
        \begin{tabular}{lrcccccccrr} 
                \hline
                {Source Name}                & Source id          &  $N_H (10^{22}$cm$^{-2})$   &  $\Gamma$      & $R$                          &Fe K$\alpha$ EW   &$\chi^2$/df    \\
                                             &                    &                           &                &                            &   (eV)          &                \\
                \hline
    1H 0323+342                   &    1    &   $<0.3$                           &  $1.85\pm 0.03$                     &  $0.16^{+0.10}_{-0.09}$    &  $34^{+12}_{-14}$      & $0.95 / 704$        &              &         &                    \\
    4U 0937-12                    &    2    &   $2.0\pm0.4$                        &  $1.80\pm 0.03$                 &  $0.15^{+0.09}_{-0.08}$      &  $79 \pm 10$      & $0.93 / 917$     &              &         &                   \\
    IGR J12415-5750               &    3    &   $1.7\pm0.7$                        &  $1.82\pm 0.05$                 &  $0.23^{+0.16}_{-0.14}$      &  $51^{+20}_{-19}$      & $0.90 / 593$        &              &         &                    \\
    IGR J16482-3036               &    4    &   $<1.6$                           &  $1.70\pm 0.17$                   &  $0.09^{+1.46}_{-0.09}$      &  $151^{+80}_{-75}$     & $0.77 / 53$         &              &         &                    \\
    IGR J18027-1455               &    5    &   $0.4^{+1.1}_{-0.4}$       &  $1.76^{+0.09}_{-0.06}$    &  $0.10^{+0.19}_{-0.10}$    &  $254^{+37}_{-40}$     & $1.08 / 535$       &              &         &                   \\
    IGR J19077-3925               &    6    &   $<1.7$                           &  $1.93^{+0.17}_{-0.15}$    &  $1.74^{+1.49}_{-0.94}$     &  $<42$      & $0.90 / 115$      &              &         &                   \\
    1RXS J211928.4+333259         &    7    &   $1.0^{+0.8}_{-1.0}$          &  $1.85^{+0.10}_{-0.11}$    &  $0.10^{+0.36}_{-0.10}$         &  $91^{+28}_{-33}$      & $1.15 / 248$      &              &         &                   \\
    Mrk 1018                      &    8    &   $1.1^{+3.9}_{-1.1}$          &  $1.87^{+0.41}_{-0.23}$    &  $0.92^{+3.54}_{-0.86}$         &  $275^{+94}_{-86}$     & $0.84 / 51$        &              &         &                   \\
    Mrk 520                       &    9    &   $0.4^{+1.3}_{-0.4}$          &  $1.57^{+0.11}_{-0.08}$    &  $0.39^{+0.30}_{-0.25}$         &  $133^{+26}_{-32}$     & $1.03 / 370$       &              &         &                   \\
    Mrk 590                       &   10   &   $0.3^{+1.5}_{-0.3}$           &  $1.65^{+0.13}_{-0.08}$    &  $0.27^{+0.35}_{-0.24}$         &  $191^{+35}_{-30}$     & $0.97 / 293$       &              &         &                   \\
    PG 1501+106                   &   11   &   $1.0^{+0.7}_{-0.6}$           &  $1.92^{+0.04}_{-0.10}$    &  $0.64^{+0.37}_{-0.23}$         &  $73^{+22}_{-23}$      & $0.92 / 433$      &              &         &                   \\
    Ark 120                       &   12   &   $<0.1$                                  &  $1.94\pm 0.01$                   &  $0.56\pm 0.06$            &  $179^{+9}_{-10}$     & $1.03 / 1289$                    &           &   &                  \\
    ESO 141-55                    &   13   &   $<0.2$                            &  $1.94^{+0.01}_{-0.02}$    &  $0.61^{+0.09}_{-0.06}$    &  $124^{+15}_{-14}$     & $0.99 / 1075$       &              &  &      \\
    ESO 33-2                      &   14   &   $1.4^{+0.8}_{-0.7}$           &  $2.23\pm 0.08$                       &  $2.62^{+0.86}_{-0.66}$      &  $72^{+23}_{-24}$      & $0.97 / 420$         &              &         &                    \\
    Fairall 1146                  &   15   &   $0.6^{+0.9}_{-0.6}$           &  $2.05\pm 0.08$                       &  $1.03^{+0.40}_{-0.36}$      &  $143^{+27}_{-25}$     & $1.20 / 417$         &              &         &                    \\
    1H 2251-179                   &   16   &   $1.3\pm 0.3$                        &  $1.75\pm 0.02$           &  $0.08\pm 0.05$            &  $8 \pm 5$       & $1.02 / 1313$    &              &         &                   \\
    IGR J14471-6414               &   17   &   $4.6\pm 2.3$                        &  $2.25^{+0.30}_{-0.25}$    &  $2.25^{+3.45}_{-1.39}$    &  $56^{+59}_{-46}$      & $1.05 / 87$      &              &         &                   \\
    IGR J14552-5133               &   18   &   $<0.7$                            &  $1.96^{+0.09}_{-0.08}$   &  $1.25^{+0.63}_{-0.45}$    &  $103^{+38}_{-39}$     & $1.07 / 211$      &              &         &                   \\
    SWIFT J1650.5+0434            &   19   &   $3.2 \pm 1.5$                     &  $1.79^{+0.14}_{-0.13}$    &  $1.04^{+0.50}_{-0.39}$    &  $55^{+27}_{-30}$      & $0.84 / 315$     &              &         &                   \\
    SWIFT J1933.9+3258            &   20   &   $1.0^{1.1}_{1.0}$                  &  $2.25^{+0.09}_{-0.15}$    &  $2.39^{+1.36}_{-0.90}$    &  $31^{+26}_{-25}$      & $0.93 / 268$     &              &         &                   \\
    IGR J19378-0617               &   21   &   $1.9 \pm 0.4$               &  $2.48 \pm 0.04$           &  $1.32^{+0.25}_{-0.22}$    &  $60 \pm 10$           & $1.01 / 742$        &              &         &                   \\
    IGR J21277+5656               &   22   &   $0.9\pm 0.2$                       &  $2.03\pm 0.03$            &  $1.77^{+0.13}_{-0.12}$    &  $58^{+5}_{-4}$      & $0.98 / 1348$    &              &         &                   \\
    MCG -06-30-015                &   23   &   $1.7^{+0.2}_{-0.1}$           &  $2.26^{+0.02}_{-0.01}$    &  $1.71^{+0.11}_{-0.09}$    &  $84 \pm 6$      & $1.15 / 1499$ &              &         &                   \\
    Mrk 1040                      &   24   &   $0.9\pm 0.3$                        &  $1.92\pm 0.02$           &  $0.66\pm 0.09$            &  $138^{+13}_{-12}$     & $1.02 / 1171$     &              &         &                   \\
    Mrk 509                       &   25   &   $<0.02$                         &  $1.75\pm 0.01$               &  $0.41\pm 0.04$            &  $69 \pm 5$      & $1.11 / 1637$     &              &         &                   \\
    Mrk 766                       &   26   &   $0.8\pm 0.3$                        &  $2.30\pm 0.03$           &  $1.47^{+0.16}_{-0.17}$    &  $40 \pm 7$      & $1.07 / 936$      &              &         &                   \\
    NGC 0985                      &   27   &   $7.1^{+1.2}_{-1.8}$           &  $2.12^{+0.08}_{-0.17}$    &  $1.90^{+1.03}_{-0.72}$    &  $30^{+34}_{-28}$      & $1.04 / 239$     &              &         &                   \\
    NGC 3516                      &   28   &   $2.9\pm 0.6$                        &  $1.73\pm 0.05$           &  $0.82^{+0.22}_{-0.19}$    &  $264^{+17}_{-18}$     & $1.08 / 710$      &              &         &                   \\
    NGC 4051                      &   29   &   $0.8\pm 0.2$                        &  $2.20\pm 0.02$           &  $2.63^{+0.21}_{-0.15}$    &  $101^{+37}_{-27}$     & $1.17 / 1267$     &              &         &                   \\
    NGC 4593                      &   30   &   $<0.3$                                  &  $1.81^{+0.02}_{-0.01}$                 &  $0.49^{+0.09}_{-0.06}$    &  $135 \pm 8$     & $1.07 / 1147$                    &             &         &                    \\
    NGC 5506                      &   31   &   $3.9\pm 0.2$                        &  $1.92\pm 0.02$           &  $1.22^{+0.11}_{-0.09}$    &  $121 \pm 7$     & $1.04 / 1328$      &              &         &                   \\
    NGC 5548                      &   32   &   $3.0\pm 0.2$                        &  $1.63\pm 0.02$           &  $0.53\pm 0.05$            &  $88 \pm 4$      & $1.05 / 1704$     &              &         &                   \\
    NGC 5995                      &   33   &   $1.9\pm 0.8$                        &  $2.02^{+0.08}_{-0.07}$    &  $1.00^{+0.39}_{-0.31}$    &  $139^{+25}_{-24}$     & $1.00 / 458$      &              &         &                   \\
    NGC 6814                      &   34   &   $0.9\pm 0.2$                        &  $1.84\pm 0.02$           &  $0.56\pm 0.07$            &  $101 \pm 5$     & $1.03 / 1413$      &              &         &                   \\
    NGC 7314                      &   35   &   $1.1\pm 0.2$                        &  $2.09^{+0.01}_{-0.02}$    &  $1.09^{+0.12}_{-0.11}$    &  $170^{+14}_{-12}$     & $1.01 / 1222$     &              &         &                   \\
    NGC 7469                      &   36   &   $<0.4$                            &  $1.89\pm 0.01$             &  $0.55^{+0.07}_{-0.06}$    &  $131^{+10}_{-9}$     & $1.07 / 1323$     &              &         &                   \\
    PG 0804+761                   &   37   &   $0.2^{+1.6}_{-0.2}$           &  $2.04^{+0.13}_{-0.10}$    &  $1.50^{+0.85}_{-0.60}$    &  $62^{+39}_{-32}$      & $0.90 / 185$     &              &         &                   \\
    SWIFT J0318.7+6828            &   38   &   $9.4^{+3.2}_{-3.1}$           &  $1.59^{+0.21}_{-0.10}$           &  $0.15^{+0.43}_{-0.15}$    &  $114^{+46}_{-32}$     & $0.96 / 169$                       &             &         &                   \\
    SWIFT J0505.7-2348            &   39   &   $15.1\pm1.4$                        &  $1.77\pm 0.10$           &  $0.55^{+0.27}_{-0.22}$    &  $53^{+21}_{-19}$      & $0.97 / 549$     &              &         &                   \\
    SWIFT J0920.8-0805            &   40   &   $8.8\pm 0.6$                        &  $1.82\pm 0.05$           &  $0.64^{+0.13}_{-0.12}$    &  $84^{+10}_{-9}$      & $0.98 / 957$     &              &         &                   \\
    IGR J09253+6929               &   41   &   $14.5\pm 1.9$                     &  $1.86\pm 0.14$             &  $0.67^{+0.60}_{-0.41}$    &  $111^{+37}_{-29}$     & $0.98 / 252$      &              &         &                   \\
    IGR J12391-1612               &   42   &   $8.0\pm 3.2$                        &  $1.79\pm 0.27$           &  $0.85^{+1.24}_{-0.71}$    &  $71^{+54}_{-51}$      & $1.00 / 111$     &              &         &                   \\
    IGR J19473+4452               &   43   &   $11.6^{+1.7}_{-2.5}$          &  $1.83^{+0.06}_{-0.17}$    &  $0.29^{+0.45}_{-0.29}$    &  $76^{+34}_{-28}$      & $1.09 / 217$     &              &         &                   \\
    IGR J20216+4359               &   44   &   $9.4^{+2.1}_{-3.5}$       &  $1.70^{+0.15}_{-0.28}$    &  $1.23^{1.22}_{-0.80}$     &  $161^{+72}_{-66}$     & $1.00 / 100$      &              &         &                   \\
    IGR J23308+7120               &   45   &   $7.7^{+4.7}_{-5.0}$           &  $1.73^{+0.30}_{-0.37}$    &  $0.78^{+1.37}_{-0.76}$    &  $<85$      & $0.97 / 57$        &              &         &                   \\
    IGR J23524+5842               &   46   &   $6.3^{+3.5}_{-3.8}$           &  $1.71^{+0.22}_{-0.24}$    &  $0.53^{+0.70}_{-0.49}$    &  $<32$   & $0.98 / 136$  &              &         &                   \\
    Mrk 348                       &   47   &   $10.9^\pm 0.7$                    &  $1.65\pm 0.05$             &  $0.27^{+0.11}_{-0.10}$    &  $58 \pm 12$      & $0.91 / 926$     &              &         &                   \\
    NGC 2110                      &   48   &   $5.1\pm 0.2$                        &  $1.67\pm 0.01$           &  $<0.14$                   &  $39^{+4}_{-5}$      & $0.97 / 1584$    &              &         &                   \\
    NGC 4258                      &   49   &   $11.4\pm 1.1$                     &  $1.68\pm 0.09$             &  $0.11^{+0.21}_{-0.11}$    &  $67^{+19}_{-20}$      & $0.99 / 567$     &              &         &                   \\
    NGC 4395                      &   50   &   $4.5\pm 1.5$                        &  $1.47^{+0.13}_{-0.12}$    &  $0.40^{+0.31}_{-0.25}$    &  $102^{+28}_{-27}$     & $0.90 / 330$      &              &         &                   \\
    NGC 5252                      &   51   &   $5.6\pm 1.2$                        &  $1.66\pm 0.09$           &  $<0.03$                   &  $68 \pm 22$      & $1.03 / 417$     &              &         &                   \\
    NGC 7172                      &   52   &   $12.6^{+0.3}_{-0.4}$          &  $1.89^{+0.01}_{-0.03}$    &  $0.62^{+0.10}_{-0.06}$    &  $61 \pm 9$      & $1.02 / 1102$ &              &         &                   \\
    
                                \hline  
        \end{tabular}
\end{table*}     
    
\addtocounter{table}{-1}         

\begin{table*}
        \centering
        \caption{Continued}
        \label{tab:best_fit}
        \begin{tabular}{lrcccccccrr} 
                \hline
                Source Name        &Source id          &  $N_H (10^{22}$cm$^{-2})$   &  $\Gamma$      & $R$                          &Fe K$\alpha$ EW   &$\chi^2$/df    \\
                                   &                   &                           &                &                            &   (eV)           &            \\
                \hline   
    ESO 103-35                    &   53   &   $24.0 \pm 1.0$                    &  $1.91\pm 0.07$                       &  $1.38^{+0.27}_{-0.23}$    &  $118^{+12}_{-14}$     & $0.97 / 841$                       &             &  &                   \\
    IGR J01528-0326               &   54   &   $18.6^{+3.0}_{-4.2}$          &  $1.78^{+0.13}_{-0.31}$    &  $2.22^{+1.68}_{-1.04}$    &  $298^{+76}_{-73}$     & $0.85 / 97$       &              &         &                   \\
    IGR J06239-6052               &   55   &   $27.4^{+2.7}_{-3.2}$          &  $2.06^{+0.19}_{-0.26}$    &  $2.45^{+1.87}_{-1.11}$    &  $<42$      & $0.84 / 178$       &              &         &                   \\
    SWIFT J1049.4+2258            &   56   &   $39.6^{+4.6}_{-4.4}$          &  $1.44^{+0.15}_{-0.18}$    &  $<0.23$                   &  $126^{+45}_{-41}$     & $0.95 / 193$      &              &         &                   \\
    IGR J14579-4308               &   57   &   $14.7^{+3.2}_{-3.9}$          &  $1.84^{+0.17}_{-0.32}$    &  $2.52^{+1.85}_{-1.24}$    &  $89^{+73}_{-48}$      & $0.87 / 96$      &              &         &                   \\
    IGR J18244-5622               &   58   &   $22.9^{+3.2}_{-3.1}$          &  $1.65\pm 0.24$                 &  $2.18^{+1.34}_{-0.85}$    &  $201^{+40}_{-39}$     & $1.05 / 220$      &              &         &                   \\
    SWIFT J1930.5+3414            &   59   &   $44.9^{+4.2}_{-4.0}$          &  $1.76$(f)                    &  $0.77^{+0.28}_{0.69}$     &  $<43$      & $1.08 / 180$     &              &         &                   \\
    Mrk 477                       &   60   &   $25.9\pm 1.7$                     &  $1.76$(f)                 &  $0.73^{+0.29}_{0.35}$     &  $59^{+30}_{-33}$      & $1.03 / 259$      &             &         &                   \\
    Mrk 6                         &   61   &   $17.7\pm 0.7$                     &  $1.76$(f)                  &  $3.18^{0.22}_{-0.24}$     &  $116 \pm 17$     & $1.03 / 1045$     &              &         &                   \\
    NGC 6300                      &   62   &   $20.7^{+0.5}_{-0.4}$          &  $1.90\pm 0.03$                 &  $1.42^{+0.16}_{-0.15}$    &  $78^{+10}_{-11}$      & $1.01 / 1100$    &              &         &                   \\
    NGC 7582                      &   63   &   $27.2\pm 1.1$                     &  $1.53\pm 0.07$            &  $1.68^{+0.25}_{-0.22}$    &  $171^{+18}_{-15}$     & $1.07 / 1133$      &             &         &                   \\
                
                                \hline  
        \end{tabular}
\end{table*}

\bibliographystyle{aa} 
\bibliography{nustaragn_v1} 

\begin{thebibliography}{38}
\expandafter\ifx\csname natexlab\endcsname\relax\def\natexlab#1{#1}\fi

\bibitem[{Akaike(1973)}]{Akaike1998}
Akaike, H. 1973, in Second International Symposium on Information Theory, ed.
  B.~N. Petrov \& F.~Csaki (Budapest: Akad\'{e}miai Kiado), 267--281

\bibitem[{{Antonucci}(1993)}]{1993ARA&A..31..473A}
{Antonucci}, R. 1993, \araa, 31, 473

\bibitem[{{Arnaud}(1996)}]{1996ASPC..101...17A}
{Arnaud}, K.~A. 1996, in Astronomical Society of the Pacific Conference Series,
  Vol. 101, Astronomical Data Analysis Software and Systems V, ed. G.~H.
  {Jacoby} \& J.~{Barnes}, 17

\bibitem[{{Asplund} {et~al.}(2009){Asplund}, {Grevesse}, {Sauval}, \&
  {Scott}}]{2009ARA&A..47..481A}
{Asplund}, M., {Grevesse}, N., {Sauval}, A.~J., \& {Scott}, P. 2009, \araa, 47,
  481

\bibitem[{{Barthelmy} {et~al.}(2005){Barthelmy}, {Barbier}, {Cummings},
  {Fenimore}, {Gehrels}, {Hullinger}, {Krimm}, {Markwardt}, {Palmer},
  {Parsons}, {Sato}, {Suzuki}, {Takahashi}, {Tashiro}, \&
  {Tueller}}]{2005SSRv..120..143B}
{Barthelmy}, S.~D., {Barbier}, L.~M., {Cummings}, J.~R., {et~al.} 2005, \ssr,
  120, 143

\bibitem[{{Bauer} {et~al.}(2015){Bauer}, {Ar{\'e}valo}, {Walton}, {Koss},
  {Puccetti}, {Gandhi}, {Stern}, {Alexander}, {Balokovi{\'c}}, {Boggs},
  {Brandt}, {Brightman}, {Christensen}, {Comastri}, {Craig}, {Del Moro},
  {Hailey}, {Harrison}, {Hickox}, {Luo}, {Markwardt}, {Marinucci}, {Matt},
  {Rigby}, {Rivers}, {Saez}, {Treister}, {Urry}, \&
  {Zhang}}]{2015ApJ...812..116B}
{Bauer}, F.~E., {Ar{\'e}valo}, P., {Walton}, D.~J., {et~al.} 2015, \apj, 812,
  116

\bibitem[{{Beloborodov}(1999)}]{1999ApJ...510L.123B}
{Beloborodov}, A.~M. 1999, \apjl, 510, L123

\bibitem[{{Brightman} \& {Ueda}(2012)}]{2012MNRAS.423..702B}
{Brightman}, M. \& {Ueda}, Y. 2012, \mnras, 423, 702

\bibitem[{{Collinge} {et~al.}(2001){Collinge}, {Brandt}, {Kaspi}, {Crenshaw},
  {Elvis}, {Kraemer}, {Reynolds}, {Sambruna}, \& {Wills}}]{2001ApJ...557....2C}
{Collinge}, M.~J., {Brandt}, W.~N., {Kaspi}, S., {et~al.} 2001, \apj, 557, 2

\bibitem[{{De Marco} {et~al.}(2013){De Marco}, {Ponti}, {Cappi}, {Dadina},
  {Uttley}, {Cackett}, {Fabian}, \& {Miniutti}}]{2013MNRAS.431.2441D}
{De Marco}, B., {Ponti}, G., {Cappi}, M., {et~al.} 2013, \mnras, 431, 2441

\bibitem[{{Del Moro} {et~al.}(2017){Del Moro}, {Alexander}, {Aird}, {Bauer},
  {Civano}, {Mullaney}, {Ballantyne}, {Brandt}, {Comastri}, {Gandhi},
  {Harrison}, {Lansbury}, {Lanz}, {Luo}, {Marchesi}, {Puccetti}, {Ricci},
  {Saez}, {Stern}, {Treister}, \& {Zappacosta}}]{2017ApJ...849...57D}
{Del Moro}, A., {Alexander}, D.~M., {Aird}, J.~A., {et~al.} 2017, \apj, 849, 57

\bibitem[{{Emmanoulopoulos} {et~al.}(2016){Emmanoulopoulos}, {Papadakis},
  {Epitropakis}, {Pech{\'a}{\v c}ek}, {Dov{\v c}iak}, \&
  {McHardy}}]{2016MNRAS.461.1642E}
{Emmanoulopoulos}, D., {Papadakis}, I.~E., {Epitropakis}, A., {et~al.} 2016,
  \mnras, 461, 1642

\bibitem[{{Esposito} \& {Walter}(2016)}]{2016A&A...590A..49E}
{Esposito}, V. \& {Walter}, R. 2016, \aap, 590, A49

\bibitem[{{Fabbiano} {et~al.}(2018){Fabbiano}, {Paggi}, {Karovska}, {Elvis},
  {Maksym}, \& {Wang}}]{2018ApJ...865...83F}
{Fabbiano}, G., {Paggi}, A., {Karovska}, M., {et~al.} 2018, \apj, 865, 83

\bibitem[{{Ghisellini} {et~al.}(1991){Ghisellini}, {George}, {Fabian}, \&
  {Done}}]{1991MNRAS.248...14G}
{Ghisellini}, G., {George}, I.~M., {Fabian}, A.~C., \& {Done}, C. 1991, \mnras,
  248, 14

\bibitem[{{Guainazzi} {et~al.}(2016){Guainazzi}, {Risaliti}, {Awaki},
  {Arevalo}, {Bauer}, {Bianchi}, {Boggs}, {Brandt}, {Brightman}, {Christensen},
  {Craig}, {Forster}, {Hailey}, {Harrison}, {Koss}, {Longinotti}, {Markwardt},
  {Marinucci}, {Matt}, {Reynolds}, {Ricci}, {Stern}, {Svoboda}, {Walton}, \&
  {Zhang}}]{2016MNRAS.460.1954G}
{Guainazzi}, M., {Risaliti}, G., {Awaki}, H., {et~al.} 2016, \mnras, 460, 1954

\bibitem[{{Harrison} {et~al.}(2013){Harrison}, {Craig}, {Christensen},
  {Hailey}, {Zhang}, {Boggs}, {Stern}, {Cook}, {Forster}, {Giommi},
  {Grefenstette}, {Kim}, {Kitaguchi}, {Koglin}, {Madsen}, {Mao}, {Miyasaka},
  {Mori}, {Perri}, {Pivovaroff}, {Puccetti}, {Rana}, {Westergaard}, {Willis},
  {Zoglauer}, {An}, {Bachetti}, {Barri{\`e}re}, {Bellm}, {Bhalerao},
  {Brejnholt}, {Fuerst}, {Liebe}, {Markwardt}, {Nynka}, {Vogel}, {Walton},
  {Wik}, {Alexander}, {Cominsky}, {Hornschemeier}, {Hornstrup}, {Kaspi},
  {Madejski}, {Matt}, {Molendi}, {Smith}, {Tomsick}, {Ajello}, {Ballantyne},
  {Balokovi{\'c}}, {Barret}, {Bauer}, {Blandford}, {Brandt}, {Brenneman},
  {Chiang}, {Chakrabarty}, {Chenevez}, {Comastri}, {Dufour}, {Elvis}, {Fabian},
  {Farrah}, {Fryer}, {Gotthelf}, {Grindlay}, {Helfand}, {Krivonos}, {Meier},
  {Miller}, {Natalucci}, {Ogle}, {Ofek}, {Ptak}, {Reynolds}, {Rigby},
  {Tagliaferri}, {Thorsett}, {Treister}, \& {Urry}}]{2013ApJ...770..103H}
{Harrison}, F.~A., {Craig}, W.~W., {Christensen}, F.~E., {et~al.} 2013, \apj,
  770, 103

\bibitem[{{Kara} {et~al.}(2016){Kara}, {Alston}, {Fabian}, {Cackett}, {Uttley},
  {Reynolds}, \& {Zoghbi}}]{2016MNRAS.462..511K}
{Kara}, E., {Alston}, W.~N., {Fabian}, A.~C., {et~al.} 2016, \mnras, 462, 511

\bibitem[{{Koss} {et~al.}(2016){Koss}, {Glidden}, {Balokovi{\'c}}, {Stern},
  {Lamperti}, {Assef}, {Bauer}, {Ballantyne}, {Boggs}, {Craig}, {Farrah},
  {F{\"u}rst}, {Gandhi}, {Gehrels}, {Hailey}, {Harrison}, {Markwardt},
  {Masini}, {Ricci}, {Treister}, {Walton}, \& {Zhang}}]{2016ApJ...824L...4K}
{Koss}, M.~J., {Glidden}, A., {Balokovi{\'c}}, M., {et~al.} 2016, \apjl, 824,
  L4

\bibitem[{{Lira} {et~al.}(2013){Lira}, {Videla}, {Wu}, {Alonso-Herrero},
  {Alexander}, \& {Ward}}]{2013ApJ...764..159L}
{Lira}, P., {Videla}, L., {Wu}, Y., {et~al.} 2013, \apj, 764, 159

\bibitem[{{Madsen} {et~al.}(2015){Madsen}, {Harrison}, {Markwardt}, {An},
  {Grefenstette}, {Bachetti}, {Miyasaka}, {Kitaguchi}, {Bhalerao}, {Boggs},
  {Christensen}, {Craig}, {Forster}, {Fuerst}, {Hailey}, {Perri}, {Puccetti},
  {Rana}, {Stern}, {Walton}, {J{\o}rgen Westergaard}, \&
  {Zhang}}]{2015ApJS..220....8M}
{Madsen}, K.~K., {Harrison}, F.~A., {Markwardt}, C.~B., {et~al.} 2015, \apjs,
  220, 8

\bibitem[{{Magdziarz} \& {Zdziarski}(1995)}]{1995MNRAS.273..837M}
{Magdziarz}, P. \& {Zdziarski}, A.~A. 1995, \mnras, 273, 837

\bibitem[{{Markowitz} {et~al.}(2014){Markowitz}, {Krumpe}, \&
  {Nikutta}}]{2014MNRAS.439.1403M}
{Markowitz}, A.~G., {Krumpe}, M., \& {Nikutta}, R. 2014, \mnras, 439, 1403

\bibitem[{{Murphy} \& {Yaqoob}(2009)}]{2009MNRAS.397.1549M}
{Murphy}, K.~D. \& {Yaqoob}, T. 2009, \mnras, 397, 1549

\bibitem[{{Nenkova} {et~al.}(2008{\natexlab{a}}){Nenkova}, {Sirocky},
  {Ivezi{\'c}}, \& {Elitzur}}]{2008ApJ...685..147N}
{Nenkova}, M., {Sirocky}, M.~M., {Ivezi{\'c}}, {\v Z}., \& {Elitzur}, M.
  2008{\natexlab{a}}, \apj, 685, 147

\bibitem[{{Nenkova} {et~al.}(2008{\natexlab{b}}){Nenkova}, {Sirocky},
  {Nikutta}, {Ivezi{\'c}}, \& {Elitzur}}]{2008ApJ...685..160N}
{Nenkova}, M., {Sirocky}, M.~M., {Nikutta}, R., {Ivezi{\'c}}, {\v Z}., \&
  {Elitzur}, M. 2008{\natexlab{b}}, \apj, 685, 160

\bibitem[{{Netzer}(2015)}]{2015ARA&A..53..365N}
{Netzer}, H. 2015, \araa, 53, 365

\bibitem[{{Poutanen} {et~al.}(2018){Poutanen}, {Veledina}, \&
  {Zdziarski}}]{2018A&A...614A..79P}
{Poutanen}, J., {Veledina}, A., \& {Zdziarski}, A.~A. 2018, \aap, 614, A79

\bibitem[{{Ramos Almeida} \& {Ricci}(2017)}]{2017NatAs...1..679R}
{Ramos Almeida}, C. \& {Ricci}, C. 2017, Nature Astronomy, 1, 679

\bibitem[{{Ricci} {et~al.}(2015){Ricci}, {Ueda}, {Koss}, {Trakhtenbrot},
  {Bauer}, \& {Gandhi}}]{2015ApJ...815L..13R}
{Ricci}, C., {Ueda}, Y., {Koss}, M.~J., {et~al.} 2015, \apjl, 815, L13

\bibitem[{{Ricci} {et~al.}(2011){Ricci}, {Walter}, {Courvoisier}, \&
  {Paltani}}]{2011A&A...532A.102R}
{Ricci}, C., {Walter}, R., {Courvoisier}, T.~J.-L., \& {Paltani}, S. 2011,
  \aap, 532, A102

\bibitem[{Sugiura(1978)}]{doi:10.1080/03610927808827599}
Sugiura, N. 1978, Communications in Statistics - Theory and Methods, 7, 13

\bibitem[{{Winkler} {et~al.}(2003){Winkler}, {Courvoisier}, {Di Cocco},
  {Gehrels}, {Gim{\'e}nez}, {Grebenev}, {Hermsen}, {Mas-Hesse}, {Lebrun},
  {Lund}, {Palumbo}, {Paul}, {Roques}, {Schnopper}, {Sch{\"o}nfelder},
  {Sunyaev}, {Teegarden}, {Ubertini}, {Vedrenne}, \&
  {Dean}}]{2003A&A...411L...1W}
{Winkler}, C., {Courvoisier}, T.~J.-L., {Di Cocco}, G., {et~al.} 2003, \aap,
  411, L1

\bibitem[{{Yaqoob}(2012)}]{2012MNRAS.423.3360Y}
{Yaqoob}, T. 2012, \mnras, 423, 3360

\bibitem[{{Zappacosta} {et~al.}(2018){Zappacosta}, {Comastri}, {Civano},
  {Puccetti}, {Fiore}, {Aird}, {Del Moro}, {Lansbury}, {Lanzuisi}, {Goulding},
  {Mullaney}, {Stern}, {Ajello}, {Alexander}, {Ballantyne}, {Bauer}, {Brandt},
  {Chen}, {Farrah}, {Harrison}, {Gandhi}, {Lanz}, {Masini}, {Marchesi},
  {Ricci}, \& {Treister}}]{2018ApJ...854...33Z}
{Zappacosta}, L., {Comastri}, A., {Civano}, F., {et~al.} 2018, \apj, 854, 33

\bibitem[{{Zdziarski} {et~al.}(2003){Zdziarski}, {Lubi{\'n}ski}, {Gilfanov}, \&
  {Revnivtsev}}]{2003MNRAS.342..355Z}
{Zdziarski}, A.~A., {Lubi{\'n}ski}, P., {Gilfanov}, M., \& {Revnivtsev}, M.
  2003, \mnras, 342, 355

\bibitem[{{Zdziarski} {et~al.}(1999){Zdziarski}, {Lubi{\'n}ski}, \&
  {Smith}}]{1999MNRAS.303L..11Z}
{Zdziarski}, A.~A., {Lubi{\'n}ski}, P., \& {Smith}, D.~A. 1999, \mnras, 303,
  L11

\bibitem[{{Zoghbi} {et~al.}(2017){Zoghbi}, {Matt}, {Miller}, {Lohfink},
  {Walton}, {Ballantyne}, {Garc{\'{\i}}a}, {Stern}, {Koss}, {Farrah},
  {Harrison}, {Boggs}, {Christensen}, {Craig}, {Hailey}, \&
  {Zhang}}]{2017ApJ...836....2Z}
{Zoghbi}, A., {Matt}, G., {Miller}, J.~M., {et~al.} 2017, \apj, 836, 2

\end{thebibliography}

\end{document}